\documentclass[prd,showpacs,showkeys,twocolumn,twoside]{revtex4}
\usepackage{amsmath,amssymb,bm,color,graphicx}

\newcommand{\lapprox}{\mathrel{\setbox0=\hbox{$<$}\raise0.6ex\copy0\kern-\wd0\lower0.65ex\hbox{$\sim$}}}
\newcommand{\be}{\begin{equation}}
\newcommand{\en}{\end{equation}}
\newcommand{\bea}{\begin{eqnarray}}
\newcommand{\eea}{\end{eqnarray}}
\newcommand*{\abs}[1]{\left|#1\right|}              
\newcommand*{\alphas}{\ensuremath{\alpha_{\mathrm{s}}}}

\newcommand*{\Nc}{\ensuremath{N_{\mathrm{c}}}}       

\newcommand*{\CP}{\mathrm{CP}}

\newcommand*{\Amp}{\ensuremath{\mathcal{A}}}
\newcommand*{\ptcl}[1]{\ensuremath{{#1}}}
\newcommand*{\B}{\ptcl{B}}
\newcommand*{\Bbar}{\ensuremath{\bar{\B}}}

\newcommand{\fBs}{f_{B_s^0}}

\begin{document}

\title{Scalar resonance effects on the $\bm {B_s-\bar{B}_s}$ mixing angle}

\author{O.~Leitner}
\email[E-mail: ]{leitner@lpnhe.in2p3.fr}
\author{J.-P. Dedonder}
\email[E-mail: ]{dedonder@univ-paris-diderot.fr}
\author{B.~Loiseau}
\email[E-mail: ]{loiseau@lpnhe.in2p3.fr}
\affiliation{Laboratoire de Physique Nucl\'eaire et de Hautes \'Energies, Groupe Th\'eorie, \\
                  Universit\'e Pierre et Marie Curie et Universit\'e Paris-Diderot, IN2P3 \& CNRS, 4 place Jussieu, 75252 Paris, France}
\author{B. El-Bennich}
\email[E-mail: ]{bennich@anl.gov}
\affiliation{Physics Division, Argonne National Laboratory, Argonne, IL 60439, USA}
\altaffiliation[Current address: ]{LFTC, Universidade Cruzeiro do Sul, Rua Galv\~ao Bueno, 868, S\~ao Paulo, 01506-000 SP, Brazil. }

\date{\today}

\begin{abstract}

The $B_s^0 \to J/\psi \phi$ and $B_s^0 \to  J/\psi f_0(980)$ decays are analyzed within generalized QCD factorization including all leading-order corrections in 
$\alphas$. We point out that the ratio of our calculated widths, $\Gamma (B_s^0\to J/\psi f_0(980), f_0(980)\to \pi^+\pi^-)/ \Gamma(B_s^0 \to J/\psi \phi,\phi\to K^+K^-)$, 
strongly indicates that $S$-wave effects in the $f_0(980)$'s daughter pions or kaons cannot be ignored in the extraction of the $B_s-\bar{B}_s$ mixing angle, 
$-2\beta_s$, from the  $B_s^0 \to \phi J/\psi$ decay amplitudes.
\end{abstract}


\maketitle

\section{Introduction}

In the Standard Model, $CP$ violation is predicted in weak decays thanks to the single phase of the Cabibbo-Kobayashi-Maskawa (CKM) matrix.
It is also well known that such a weak phase is not sufficient to generate a $CP$ violating decay amplitude. Strong phases are necessary and their 
strength may significantly enhance the effect of the weak phase. Therefore, hadronic effects, such as resonances of daughter particles
in $S$- and higher waves, require a careful analysis in the determination of $CP$ violating phases in hadronic two- and three-body
decays~\cite{Leitner:2010ai,ElBennich:2006yi,Boito:2008zk,ElBennich:2009da}.

The antimatter-matter asymmetry is expected to be very small in weak decays of $B_s$ mesons; any observed deviation may well be a signal of  
physics whose origins lie beyond the Standard Model. In the  $B_s^0 \to   J/\psi \phi $ channel, recent measurements by the CDF~\cite{Aaltonen:2007he} 
and  D$\emptyset$~\cite{Abazov:2008fj,Abazov:2008jz} Collaborations of the $B_s-\bar{B}_s$ mixing phase, $-2 \beta_s$, while not definitive, are 
considerably larger than Standard Model predictions. Taking advantage of the fact that the $B_s^0 \to  J/\psi f_0(980)$ channel does not  require any 
angular analysis, one can compute the ratio between the  $B_s^0 \to J/\psi \phi$ and $B_s^0 \to J/\psi f_0(980)$ decay widths in order to estimate the 
$\pi^+ \pi^-$ $S$-wave effect on the value of $\beta_s$. A first qualitative attempt to predict the ratio, 
\begin{eqnarray}
\mathcal{R}_{f_0/\phi}=\frac{\Gamma (B_s^0 \to  J/\psi f_0(980),  f_0(980) \to \pi^+ \pi^-)}{\Gamma(B_s^0 \to  J/\psi \phi, \phi \to K^+ K^-)}\ ,
\end{eqnarray}
was made by Stone and Zhang~\cite{Stone:2008ak} and gives a result of the order of  $20\%-30\%$. 
Their estimate relies on experimental data on $D_s^+ \to f_0(980) \pi^+$ and $D_s^+ \to \phi \pi^+$ decays and seems to indicate that the $S$-wave 
contribution of $f_0(980)\to K^+K^-$ cannot be ignored when analyzing the angle $\beta_s$ in $B_s^0 \to J/\psi \phi$.  Likewise, Xie {\em et al\/}. found 
the effect  of an $S$-wave component on $2\beta_s$ to be of the order of $10\%$ in the $\phi$ resonance region~\cite{Xie:2009fs}.  

Based on the QCD factorization (QCDF) formalism we perform a first robust calculation of the ratio $\mathcal{R}_{f_0/\phi}$. To this end, 
all the available observables (polarizations and branching ratio in $B_s^0 \to J/\psi \phi$) are used to effectively constrain the analysis of the $B_s^0
\to J/\psi \phi$ channel. The branching ratio and $CP$ asymmetry are then predicted for $B_s^0 \to J/\psi f_0(980)$, where we assume that merely the 
$s \bar s$ component  of the $f_0(980)$ is involved in the hadronic $B_s \to f_0(980)$ transition matrix element. 

In Section~\ref{sectwo} we introduce the general expressions for the $B_s^0 \to J/\psi \phi$ and $B_s^0 \to J/\psi f_0(980)$ weak decay amplitudes 
whereas Sections~\ref{secthree} and~\ref{secfour} provide the details on the leading order corrections in $\alphas$ for both these amplitudes, respectively.
In Section~\ref{secfive}, we list all numerical values of input parameters and briefly recall our model for the $B_s \to f_0(980)$ transition form 
factor~\cite{ElBennich:2008xy} on which the ratio $\mathcal{R}_{f_0/\phi}$ directly depends; we also define the parametrization for the $B_s\to \phi$ 
form factor. Section~\ref{secsix} is devoted to our results and, finally, conclusions are drawn in Section~\ref{secseven}.

\section{General form of the $\bm{\B_s^0 \to \phi J/\psi }$ and $\bm{\B_s^0 \to f_0(980) J/\psi}$ decay amplitudes\label{sectwo} }

It is important to realize beforehand that the application of QCDF, following Refs.~\cite{Beneke:2000ry,Beneke:2001ev,Beneke:2003zv,Beneke:2006hg}, 
to $B_s^0$ decays into a heavy-light final state is  not self-evident. In both final states, $\phi J/\psi$ and  $f_0(980) J/\psi $, the $s$-spectator quark
is absorbed by the  light  meson while the emitted meson is heavy, in which case QCDF is not reliable~\cite{Beneke:2000ry}. Nonetheless, as argued 
in Refs.~\cite{Chay:2000xn,Cheng:2001ez} and more recently in Ref.~\cite{Beneke:2008pi}, the production of a heavy charmonium $\bar qq$ pair 
bears ``color transparency" properties similar to those of a light meson, provided this color-singlet pair is small compared to the inverse strong 
interaction scale, $1/\Lambda_\mathrm{QCD}$. This was explicitly demonstrated in next-to-leading order calculations for exclusive $B$ decays to 
$J/\psi$ final states ($J/\psi K, J/\psi K^*$), where infrared divergences were shown to cancel~\cite{Chay:2000xn,Cheng:2001ez}.

In the following, we present the $B_s^0$ decay amplitudes in which the short- and long-distance contributions are factorized in the approximation of a 
quasi two-body state, $M_1 M_2$, where either $M_1 M_2 = f_0(980) J/\psi$ or $M_1 M_2 = \phi J/\psi$. We begin with the $B_s^0\to \phi J/\psi$ amplitude 
which can be written for each helicity, $h=-1,0,1$, as~\cite{Beneke:2006hg},
\begin{widetext}
\begin{eqnarray}\label{AVV}
   {\cal A}_{\B_s^0\to \phi J/\psi}^h  =  \sum_{q=u,c} \lambda_q  \Bigg\{A^h_{\phi J/\psi} \biggl[ \delta_{qc}\,\Bigl(a_2^{q,h}(m_b) 
   + \  \zeta^h  \Bigr) + a_3^{q,h}(m_b)+  a_5^{q,h}(m_b) + Êa_7^{q,h}(m_b) + a_9^{q,h}(m_b) 
\biggr] \Bigg\}_{\phi J/\psi} \hspace*{-2mm}.
\end{eqnarray}
\noindent
Summing over all the possible helicities, the squared modulus of the total amplitude reads 
\begin{eqnarray} \label{ampvv}
\bigl|{\cal A}_{\B_s^0 \to \phi J/\psi}\bigr|^2 & = & \bigl|{\cal A}_{\B_s^0 \to \phi J/\psi}^{h=-1}\bigr|^2 + \bigl|{\cal A}_{\B_s^0 \to \phi J/\psi}^{h=0}\bigr|^2    
 +  \bigl|{\cal A}_{\B_s^0 \to \phi J/\psi}^{h=+1}\bigr|^2 . 
\end{eqnarray}
The $\bar B_s^0 \to \phi J/\psi$ decay amplitude is obtained by exchange of helicity signs, $h=+1 \to h=-1$, and replacing $\lambda_q$ by its complex conjugate. 
The $\B_s^0 \to f_0(980) J/\psi$ amplitude is,
\begin{eqnarray}\label{ASV}
   {\cal A}_{\B_s^0 \to f_0 J/\psi} & = &   \sum_{q=u,c} \lambda_q  \Bigg\{A_{f_0 J/\psi} \biggl[ \delta_{qc}\,\Bigl(a_2^q(m_b) + \zeta  \Bigr) 
    + a_3^{q}(m_b)+ a_5^{q}(m_b)  + \  a_7^{q}(m_b)  + a_9^{q}(m_b)   \biggr] \Bigg\}_{ f_0 J/\psi}  \hspace*{-2mm}.
\end{eqnarray}
\end{widetext}
The different elements entering in the amplitudes~(\ref{AVV}) and (\ref{ASV}) are defined in Eqs.~(\ref{lambda}), (\ref{deffavs}), (\ref{AhM1M2}), (\ref{defa}) and (\ref{bdef}).
The $CP$ conjugate $\bar B_s^0$ decay amplitude is again found by replacing $\lambda_q$ by its complex conjugate.

With the generic amplitude, ${\cal A}_{\B_s^0 \to M_1 J/\psi}$, the branching ratio,
\begin{multline}\label{br}
\lefteqn{ {\cal B}(\B_s^0 \to M_1 J/\psi) =  \frac{1}{\Gamma_{B_s^0}}\frac{1}{16 \pi m_{B_s^0}} \,} \\
 \times \lambda^{1/2}\Bigl(1,m_{M_1}^2/m_{B_s^0}^2,m_{J/\psi}^2/m_{B_s^0}^2 \Bigr)\,  
\bigl |{\cal A}_{\B_s^0 \to M_1 J/\psi}\bigr |^2\ ,
\end{multline}
can be computed.  The $J/\psi$ mass is noted $m_{J/\psi}$ while  $m_{M_1} = m_{f_0(980)}$ or $m_{\phi}$ denote the $f_0(980)$ and $\phi$ masses;  
the triangle function is $\lambda(x,y,z)=(x+y-z)^2-4 xy$. In 
Eq.~(\ref{br}), 
$\Gamma_{B_s^0}=1/\tau_{B_s^0}$ is the $B_s^0$ decay width with  $\tau_{B_s^0}=(1.470 \pm 0.026)$ ps~\cite{Amsler:2008zzb} and  
$m_{B_s^0}$ is the $B_s^0$ mass.  For the CKM elements in Eqs.~(\ref{AVV}) and~(\ref{ASV}) we use the Wolfenstein parametrization,  
\begin{align}
\label{lambda}
\lambda_u & = V_{ub}^{\star}V_{us}= A \lambda^4 \left(\rho + i \eta \right)\ , \nonumber \\
\lambda_c & = V_{cb}^{\star}V_{cs}= A \lambda^2 \left(1-\frac{\lambda^2}{2} \right)  \ ,
\end{align}
with the Wolfenstein parameters $A=0.814$, $\rho=0.1385$, $\eta=0.358$ and $\lambda=0.2257$~\cite{Amsler:2008zzb}.

\subsection{Non-perturbative amplitude}

\subsubsection{The case of the scalar-vector decay}

The scalar-vector factor, $A_{f_0 J/\psi}$, in Eq.~\eqref{ASV} is given by,
\begin{multline}\label{deffavs} 
A_{f_0 J/\psi} = \langle f_0(p_{f_0})| \bar b\, \gamma_{\mu}(1-\gamma_5) s | B_s^0(p_{B_s^0})\rangle  \\
\times \langle J/\psi(p_{J/\psi},\varepsilon_{J/\psi}^{\ast})|\bar{c}\gamma^{\mu}c|0 \rangle\ ,
\end{multline}
where the hadronic matrix element  which describes the transition between the $B_s^0$ and a scalar meson, $f_0$, with the respective four-momenta
  $p_{B_s^0}$ and $p_{f_0}$ is~\cite{Melikhov:2001zv},  
\begin{multline}\label{defbs}
\langle f_0(p_{f_0})| \bar b\, \gamma_{\mu}(1-\gamma_5) s | B_s^0(p_{B_s^0})  \rangle = \\
\Bigl( p_{B_s^0} + p_{f_0}  - \frac{m_{B_s^0}^{2}-m_{f_0}^{2}}{q^{2}}q \Bigr)_{\!\!\mu} F_1^{B_s^0\to f_0}(q^{2})  \\
 + \, \frac{m_{B_s^0}^{2}-m_{f_0}^{2}}{q^{2}} q_{\mu}\ F_{0}^{B_s^0\to f_0}(q^{2})\ , \hspace{2mm}
\end{multline}
with $q=p_{B_s^0}-p_{f_0}$, $q^{2}=m_{J/\psi}^2$ and where $F_{1}^{B_s^0\to f_0}(q^{2}) $ and $F_{0}^{B_s^0\to f_0}(q^{2})$ are the vector and scalar form 
factors, respectively. In Eq.~\eqref{deffavs}, the leptonic decay constant, $f_{J/\psi}$, of the $J/\psi$ vector meson, with  four-momentum, $p_{J/\psi}$, and 
polarisation, $\varepsilon_{J/\psi}^{\ast}$, is defined as,
\begin{equation}\label{deffv}
\langle J/\psi(p_{J/\psi},\varepsilon_{J/\psi}^{\ast})|\bar{c}\gamma^{\mu}c|0
\rangle = \\
-i f_{J/\psi}m_{J/\psi}\varepsilon_{J/\psi}^{\mu \ast}\ .
\end{equation}

The scalar-vector factor, given by the product of Eqs.~\eqref{defbs} and \eqref{deffv}, is then obtained as, 
\begin{equation}\label{defavs} 
   \hspace*{-1mm} A_{f_0 J/\psi} = 
-i \frac{G_F}{\sqrt2}  2m_{J\!/\psi}\,\epsilon_{J\!/\!\psi}^* \cdot p_{B_s^0} \,F_1^{B_s^0\to f_0}(m_{J\!/\!\psi}^2) f_{J/\psi} \, , 
\end{equation}
with $4 m_{J/\psi}^2 \big | \epsilon_{J/\psi}^* \cdot p_{B_s^0} \big |^2= m_{B_s^0}^2 \, \lambda^{1/2}(m^2_{B_s^0}, m^2_{J/\psi},m^2_{f_0})$ and
the Fermi constant,  $G_F=1.16 \times 10^{-5} {\rm GeV}^{-2}$.  The $B_s^0\to f_0$ transition form factor $F_1^{B_s^0\to f_0}(m_{J/\psi}^2)$ will be 
discussed in Section~\ref{secfive}.

\subsubsection{The case of the vector-vector decay}

For the case of two vector mesons, $M_1$ and $ M_2$, the helicity formalism requires the introduction of three polarization four-vectors, ${\epsilon}_{M_j,k}$ 
$(j=1,2$ and $k=1,2,3)$ for each spin-1 particle, $M_{j}$,
\begin{align}\label{aeq1}
{\epsilon}_{M_j,1}&= (0, \vec {\epsilon}_{M_j,1})\ , \nonumber \\
{\epsilon}_{M_j,2} &= (0, \vec{\epsilon}_{M_j,2})\ , \nonumber \\
{\epsilon}_{M_j,3} &={\left(|{\vec p_{M_j}}|/m_{M_j} , E_{M_j} {\hat {p}_{M_j}}/m_{M_j} \right)}\ .
\end{align}
where $m_{M_j}$, $p_{M_j}$ and $E_{M_j}$ are  the mass, the momentum and the energy of the vector meson, $M_{j}$, respectively. The 
energies $E_{M_1}, E_{M_2}$ are given by,
\begin{equation}\label{aeq2}
 E_{M_{1,2}}=\frac{1}{2 m_{M_{2,1}}}\Bigl(m^2_{B_s^0}-m^2_{M_1}-m^2_{M_2}   \Bigr)\ .
\end{equation}
In Eq.~\eqref{aeq1}, ${\hat {p}_{M_j}}$ is defined as the unit vector along the momentum:  ${\hat {p}_{M_j}}={\vec p_{M_j}}/|{\vec p_{M_j}}|$.

The three polarization four-vectors, ${\epsilon}_{M_j,k}$, also satisfy the following relations,
\begin{equation}
 {{\epsilon}_{M_j,k}}^2 = -1\ ,  \; {\rm and} \;\; {\epsilon}_{M_j,k} \cdot {\epsilon}_{M_j,l} = 0\ , \ \
 \mathrm{for} \  k \neq l\ .
\end{equation}
The vectors $\vec {\epsilon}_{M_j,1}$, $\vec {\epsilon}_{M_j,2}$ and $\vec {\epsilon}_{M_j,3}$ form an orthogonal basis in which $\vec {\epsilon}_{M_j,1}$ and 
$\vec {\epsilon}_{M_j,2}$ describe the  transverse polarizations while  $\vec {\epsilon}_{M_j,3}$ is the  longitudinal polarization vector. 
With these three vectors one builds up the helicity basis,
\begin{align}\label{defeps}
\epsilon_{M_j,+} &= \frac{1}{\sqrt{2}}\left({\epsilon}_{M_j,1} + i \; {\epsilon}_{M_j,2} \right)= \frac{1}{\sqrt{2}}(0,+1,i,0)\ ,  \nonumber \\ 
\epsilon_{M_j,-} &=  \frac{1}{\sqrt{2}}\left({\epsilon}_{M_j,1} - i \; {\epsilon}_{M_j,2} \right)= \frac{1}{\sqrt{2}}(0,-1,i,0)\ , \nonumber \\
\epsilon_{M_j,0} &= {\epsilon}_{M_j,3}\ .
\end{align}
and $\epsilon_{M_1,\pm}=\epsilon_{M_2,\mp}$. In Eq.~\eqref{defeps}, the new four-vectors $\epsilon_{M_j,+}, \epsilon_{M_j,-}$ and $\epsilon_{M_j,0}$ are  eigenvectors 
of the helicity operator  corresponding  to the eigenvalues  $h= +1, -1$ and $0$, respectively.

The vector-vector factor, $A_{M_1 M_2}^h$,  in Eq.~\eqref{AVV} is
\begin{eqnarray}
\label{AhM1M2}
A_{M_1 M_2}^h & = & \langle M_{1}(p_{M_1},\varepsilon_{M_1}^{\ast})|\bar b\, \gamma_{\mu}(1-\gamma_5) q |B_s^0 (p_{B_s^0})\rangle  \nonumber \\ 
  &   & \hspace{5mm} \times \ \langle M_{2}(p_{M_2},\varepsilon_{M_2}^{\ast})|\bar{q}\gamma^{\mu}q'|0\rangle\ ,
\end{eqnarray}
where, in the $B^0_s$ rest-frame, the vector mesons $M_{1}$ and $M_{2}$ have opposite momentum $\vec {p}_{M_1}=  -\vec {p}_{M_2}$ along 
the $z$-direction and $\epsilon_{M_j,0} \cdot p_{M_j}=0$. 

The matrix hadronic element of a $P\to V$ transition can be decomposed into Lorentz invariants as~\cite{Melikhov:2001zv,Cheng:2001ez,Li:2003hea}
\begin{widetext}
\begin{eqnarray}
\label{defvv}
\langle M_{j}(p_{M_j},\varepsilon_{M_j}^{\ast})| \bar b \, \gamma_{\mu}(1-\gamma_5) q  |B_s^0 (p_{B_s^0})\rangle & = &
\varepsilon_{M_j,\mu}^\ast(m_{B_s^0}+m_{M_j})A_1^{B_s^0 \to M_j}(q^2) -  (p_{B_s^0}+p_{M_j})_{\mu}({\varepsilon_{M_j}^\ast}\cdot{p_{B_s^0}})
\frac{A_2^{B_s^0 \to M_j}(q^2)}{m_{B_s^0}+m_{M_j}} \nonumber   \\
 &  &  \hspace*{-4cm}  - \ q_{\mu}({\varepsilon_{M_j}^\ast}\cdot{p_{B_s^0}})\frac{2m_{M_j}}{q^2}\Bigl[A_3^{B_s^0 \to M_j}(q^2)-A_0^{B_s^0 \to M_j}(q^2)\Bigr] 
    + i\epsilon_{\mu\nu\alpha\beta}\, \varepsilon_{M_j}^{\ast\nu}p_{B_s^0}^{\alpha}p^{\beta}_{M_j} \frac{2V^{B_s^0 \to M_j}(q^2)}{m_{B_s^0}+m_{M_j}}\ ,
\end{eqnarray}
\end{widetext}
where the form factors $A_0^{B_s^0 \to M_j}(q^2)$, $A_1^{B_s^0 \to M_j}(q^2)$, $A_2^{B_s^0 \to M_j}(q^2)$ and $A_3^{B_s^0 \to M_j}(q^2)$ obey the 
following exact relations,
\begin{eqnarray}
  A_3^{B_s^0 \to M_j}(q^2) & = &  \frac{m_{B_s^0}+m_{M_j}}{2m_{M_j}} A_1^{B_s^0 \to M_j}(q^2) \nonumber \\  
   & - & \frac{m_{B_s^0}-m_{M_j}}{2m_{M_j}}A_2^{B_s^0 \to M_j}(q^2)\ ,  
\end{eqnarray}
as well as for $q^2=0$, $A_3^{B_s^0 \to M_j}(0) =A_0^{B_s^0 \to M_j}(0)$.

Specifically for $M_1=\phi$, and $M_2=J/\psi$, the helicity dependent vector-vector factor $A_{\phi J/\psi}^h$ in Eq.~\eqref{AVV} has thus the following form,
\begin{subequations}
\begin{eqnarray}
\label{defavv1}
A_{\phi J/\psi}^{(h=0)} & = & i\,\frac{G_F}{\sqrt2} f_{J/\psi} \Biggl[ -m_{\phi}(m_{B_s^0}+ m_{\phi}) A_1^{B_s^0\to \phi}(m_{J/\psi}^2) \nonumber \\
 & & \hspace*{-1cm} + \ \bigl(m_{B_s^0}^2+ m_{\phi}^2-m_{J/\psi}^2\bigr) A_0^{B_s^0\to \phi}(m_{J/\psi}^2)\Biggl]\ ;  
\end{eqnarray} 
\begin{eqnarray}
\label{defavv2}
A_{\phi J/\psi}^{(h=\pm 1)} & = & i\,\frac{G_F}{\sqrt2} m_{B_s^0} m_{J/\psi} f_{J/\psi} F_{\mp}^{B_s^0\to \phi}(m_{J/\psi}^2)\, . \hspace{5mm}
\end{eqnarray}
\end{subequations}
 In Eq.~\eqref{defavv2}, the transition form factors $F_{\pm}^{\B_s^0\to \phi}(q^2=m_{J/\psi}^2)$ are 
\begin{eqnarray}
\label{deffpm}
   F_{\pm}^{B_s^0\to \phi}(m_{J/\psi}^2)  & = & \left  (1+\frac{m_{\phi}}{m_{B_s^0}} \right ) A_1^{B_s^0\to \phi}(m_{J/\psi}^2) \nonumber \\
   &  \mp & \frac{2 |\vec p_{B_s^0}| }{m_{B_s^0}+m_{\phi}} V^{B_s^0\to \phi}(m_{J/\psi}^2)\ ,
\end{eqnarray}
where the center-of-mass momentum $|\vec p_{B_s^0}|$ is defined as,
\begin{equation}
|\vec p_{B_s^0}| =  \frac{\sqrt{\left(m_{B_s^0}^2- M_+^2 \right) \left(m_{B_s^0}^2 - M_-^2  \right)}}{2 m_{B_s^0}}\ ,
\end{equation}
with $M_\pm =m_{J/\psi} \pm m_\phi$. We note that a somewhat different form for $A_{\phi J/\psi}^{(h=0)}$ was derived in Ref.~\cite{Li:2003hea}, which
seems to approximate the vector mesons as light mesons.
The form factors $A_0^{B_s^0\to \phi}(m_{J/\psi}^2)$ and $A_1^{B_s^0\to \phi}(m_{J/\psi}^2)$ in Eqs.~\eqref{defavv1}  and~\eqref{deffpm}, as well as  
$V^{B_s^0\to \phi}(m_{J/\psi}^2)$ in Eq.~\eqref{deffpm} are defined in Section~\ref{secfive}. Ref.~\cite{Beneke:2006hg} asserts that when neglecting vector 
meson masses,  Eq.~\eqref{defavv1} reduces to,
\begin{eqnarray} 
A_{\phi J/\psi}^{(h=0)} & = & i\,\frac{G_F}{\sqrt2} f_{J/\psi} m_{B_s^0}^2 A_0^{B_s^0\to \phi}(m_{J/\psi}^2) \ . 
\end{eqnarray}
The numerical effects in the calculated values of $B_s^0 \to J/\psi \phi$ and $B_s^0 \to J/\psi f_0(980)$ branching ratios are too important to justify such an approximation.

\subsection{Perturbative amplitude}

The  $a_n^{q,h}(\mu)$ coefficients that appear in Eqs.~(\ref{AVV}) and~(\ref{ASV}) are linear combinations of Wilson coefficients, $C_n(\mu)$,
either at the scale $\mu =m_b$ or $m_b/2$ (see below):
\begin{eqnarray}\label{defa}
 \lefteqn{a_n^{q,h}(m_b)  = \left [C_n(m_b) + \frac{C_{n\pm 1}(m_b)}{N_c} \right ] N_n(J/\psi) } \nonumber \\ 
   & + & P_n^{q,h}(J/\psi) +  \frac{C_{n\pm 1}(m_b)}{N_c} \frac{C_F\ }{4\pi}  \alpha_s(m_b)  V_n^h(J/\psi) \nonumber  \\ 
   & + &   \pi C_F \alpha_s(m_b/2) \frac{C_{n\pm 1}(m_b/2)}{N_c^2} H_n^h(M_1 J/\psi)\ .
\end{eqnarray} 
The superscript, $(h)$, explicits the helicity dependence of $a_{n}^{q,h}(\mu)$ in the case where $B_s^0$ decays into two vector mesons. This superscript is 
dropped in the scalar-vector case. There is no flavor dependence in $a_{n}^{q,h}(\mu)$ for $n=1,2$. In Eq.~(\ref{defa}), the upper (lower) signs in $C_{n\pm 1}(\mu)$ 
apply when $n$ is odd (even) and
\begin{equation}
   N_n (J/\psi) = 0,\  n\in\{6,8\}, \  \mathrm{else}  \ \ N_n (J/\psi) =1\, .
\end{equation}
The Wilson coefficients, $C_n(\mu)$, in the Naive Dimensional Regularization (NDR) scheme are taken at the hard scale $m_b$ for the vertex, 
$V_n^h(J/\psi)$, and penguin, $P_n^{q,h}(J/\psi)$, corrections, whereas in the hard scattering, $H_n^h(M_1 J/\psi)$, amplitudes they are 
evaluated at $m_b/2$ since those contributions involve the spectator quark. The strong coupling constants at these scales are $\alphas(m_b)=0.224$ 
and $\alphas(m_b/2)=0.286$~\cite{Amsler:2008zzb}, while the number of active flavors is $n_F=5$, the color number $N_c=3$ and  $C_F=(N_c^2-1)/2 N_c$.

\subsection{Suppressed higher order corrections and possibility of new physics}

There are no contributions, such as given by the annihilation operators derived in Ref.~\cite{Beneke:2003zv},  to the two decays considered here.
This is because for the final states, $J/\psi\phi$ and $J/\psi f_0(980)$, both mesons are simultaneously
flavor \textit{and} color singlets. At tree level, for instance, the $W^\pm$ exchange diagram produces the charmonium $\bar cc$, yet the creation
of the $\bar ss$ which hadronizes to an $f_0(980)$ or $\phi$ must proceed via multiple gluons or by means of photon/$Z$ exchange.
The annihilation is thus either strongly (Zweig) suppressed in $\alpha_s$ or the suppression is in the electromagnetic coupling constant
 $\alpha_{\mathrm{em}}$.
 
On the other hand, as will be discussed in Section~\ref{secsix}, if we account for vertex, penguin and hard scattering corrections only, the $B_s^0\to J/\psi\phi$
observables are only moderately well reproduced. As can be seen in Table~\ref{sumfit}, the branching ratio, for instance, is about 20\% too large 
(although still within the experimental errors). We therefore allow for additional phenomenological amplitudes that mock up ``other" contributions, 
be it from annihilation topologies expected to be strongly suppressed or possible physics beyond the Standard Model~\cite{Chiang:2009ev}. 
These are included in Eqs.~\eqref{AVV} and \eqref{ASV} with the amplitudes, $\zeta^h$ and $\zeta$, conveniently scaled as, 
\begin{equation}\label{bdef}
\zeta^{(h)}  =\frac{B_{M_1 J/\psi} }{A_{M_1 J/\psi}^{(h)}} \, X_C   \ .
\end{equation}
The factor $B_{M_1 J/\psi}$ is chosen to be a product of decay constants, either
\begin{equation}\label{defbvs}
  B_{f_0 J/\psi} =  - i\,\frac{G_F}{\sqrt{2}}\,\fBs\, \bar f_{f_0}\, f_{J/\psi}\ ,
\end{equation}
if $M_1=f_0(980)$ or
\begin{equation}\label{defbvv}
  B_{\phi J/\psi} =  i\,\frac{G_F}{\sqrt{2}}\,\fBs\,  f_{\phi}\,  f_{J/\psi}\ ,
\end{equation}
if $M_1=\phi$, while the factor $X_C$ is a complex parameter discussed in Section~\ref{modelparam}.  We note that the decay constant, $f_{f_0}$, vanishes 
due to charge conjugation invariance, wherefore the scalar light cone  distributions amplitude (LCDA) is normalized to  
$\bar f_{f_0} =   f_{f_0} m_{f_0}/ (m_{u,d}(\mu) - m_{u,d}(\mu))$, which is finite~\cite{Cheng:2005nb}. We shall return to this issue in Section~\ref{secfour}.

\subsection{The ratio $\bm{\mathcal{R}_{f_0/\phi}}$}

Prior to discussing the various $\alpha_s(\mu)$ corrections to the amplitudes, $a_{n}^ {p,h}(\mu)$, it may be of interest to observe the qualitative behavior 
of the ratio, $\mathcal{R}_{f_0/\phi}$, in terms of the scales $\Lambda_\mathrm{QCD}$ and $m_b$. A naive factorization analysis yields a hierarchy of helicity 
amplitudes for $B$ into  vector-vector decays~\cite{Beneke:2006hg},
\begin{eqnarray}
 \lefteqn{\hspace*{-8mm} \mathcal{A}^{(h=0)}_{B_s^0 \to \phi J/\psi} :  \mathcal{ A}^{(h=+1)}_{B_s^0 \to \phi J/\psi}  :   \mathcal{ A}^{(h=-1)}_{B_s^0 \to \phi J/\psi} }  
   \hspace*{1cm} \nonumber \\
   &  \Longleftrightarrow & \   1 :  \frac{\Lambda_\mathrm{QCD}}{m_b}  :  \left ( \frac{\Lambda_\mathrm{QCD}}{m_b} \right )^{\! 2}  ,
 \label{helicityratio}
\end{eqnarray}
while for $\bar B_s$ mesons the signs are exchanged ($h=+ 1 \to h=-1$). Furthermore, the amplitudes $\mathcal{A}^{(h=0)}_{B_s^0 \to \phi J/\psi}$
and  $\mathcal{A}_{B_s^0 \to f_0 J/\psi}$ are of same order in $\Lambda_\mathrm{QCD}/m_b$. With this estimation, the ratio $\mathcal R_{f_0/\phi}$ 
we are interested in becomes,
\begin{eqnarray}
  \mathcal{R}_{f_0/\phi} & =& \frac{\big |\mathcal{A}_{B_s^0 \to f_0 J/\psi} \big |^2} { \big | \mathcal{A}^{(h=0)}_{B_s^0 \to \phi J/\psi}\big |^2 + 
  \big | \mathcal{A}^{(h=-1)}_{B_s^0 \to \phi J/\psi} \big |^2 +
  \big | \mathcal{A}^{(h=+1)}_{B_s^0 \to \phi J/\psi} \big |^2} \nonumber \\
  & \simeq &  \mathcal{O}(1) +  \mathcal{O} \left(\frac{\Lambda_\mathrm{QCD}}{m_b}\right)^{\!\!2} + \mathcal{O} \left ( \frac{\Lambda_\mathrm{QCD}}{m_b} \right )^{\!\!4}  .
\end{eqnarray}
Hence, $R_{f_0/\phi}$ is $\mathcal{O}(1)$ for $\Lambda_\mathrm{QCD}/m_b$ corrections. 

Nonetheless, non-perturbative {\rm hadronic} effects can spoil the naive factorization and violate the hierarchy in Eq.~\eqref{helicityratio}; 
so do electromagnetic penguin contributions where a photon with small virtuality subsequently converts into a vector meson~\cite{Beneke:2005we}.

\section{QCDF corrections for $\bm{B_s^0 \to \phi J/\psi}$ decay amplitudes \label{secthree}}

Due to the structure of the four-quark operators in heavy quark effective theory and the conservation of the flavor quantum numbers, the 
final state $M_1 M_2 = \phi J/\psi$ is created from the transition $B_s^0 \to \phi$ and the production of  $J/\psi$  from vacuum. As  
discussed in Section~\ref{sectwo}, the decay amplitudes at leading order in $\Lambda_{\mathrm{QCD}}/m_b$ and $\alphas(m_b)$ are given 
by the factorized product of a transition form factor and a decay constant. Following Ref.~\cite{Beneke:2006hg}, we only give QCD corrections 
that explicitly appear in the amplitude ${\cal A}_{\B_s^0\to \phi J/\psi}^h$ of Eq.~\eqref{AVV}.

We discard terms proportional to $r = (m_{J/\psi}/m_{B_s} )^2 \simeq 1/3$ in vertex corrections which stem from the presence of the charm 
quark in the loop diagram; we have numerically checked that their contributions to the $a_n^{q,h}\!(\mu)$ coefficients are negligible, all the more 
so when seen in the light of the large hadronic uncertainties of  the form factors  [see Sections~\eqref{bvform} and \eqref{bsform}]. We note that 
in the limit $r \to 0$, one recovers the vertex correction known from, for example, $B \to \pi\pi$ which is of course infrared safe.

Since the coefficients in the Gegenbauer expansion of the LCDA are poorly known for the scalar mesons, and only  with non-negligible errors 
for the vector mesons $V= \phi$ and $V= J/\psi$, we limit ourselves to leading terms in the expansion. The leading twist-2 
distribution and twist-3 two particle distribution amplitudes are approximated by
\begin{equation}\label{lt2}
\phi_V (x)  = 6 x (1-x) 
\end{equation}
and 
\be\label{lt3}
 \varphi_V (x) = 3 (2 x -1)\ , 
\en
respectively. In the annihilation and hard scattering amplitudes, the chiral coefficient, $r_{\chi}^{V}(\mu)$, is defined as 
\begin{equation}\label{defrchiv}
  r_{\chi}^{V}(\mu)= \frac{2 m_{V}}{m_b(\mu)}\frac{f_{V}^{\perp}(\mu)}{f_{V}} \simeq \frac{2 m_{V}}{m_b(\mu)}\ ,
\end{equation}
where $f_{V}^{\perp}(\mu)$ is the transverse decay constant for any vector $V$ and $\mu=m_b/2$.

\subsection{Penguin contributions}

The penguin contributions to the amplitude in Eq.~(\ref{AVV}) stems from the positive helicity, $h=+ 1$, amplitudes $P_{7,9}^{q,h=+ 1}(J/\psi)$ 
given in Ref.~\cite{Beneke:2006hg},
\begin{multline}\label{vvp7hpm}
  \lefteqn{ P_{7,9}^{q,h=+ 1}(J/\psi)\, = \,  -\frac{\alpha_e}{3\pi}  C^{\mathrm{eff}}_{7\gamma}(\mu)\,\frac{m_{B_s^0} m_b}{m_{J/\psi}^2}\, +    \frac{2 \alpha_e}{27\pi}} 
 \\  
\times   \Bigl(C_1(\mu)+\Nc C_2(\mu)\Bigr) \Biggl[\delta_{qc} \ln\frac{m_c^2}{\mu^2} + \delta_{qu} \ln\frac{\nu^2}{\mu^2} + 1\Biggr] \ ,
  \end{multline}
whereas $P_{7,9}^{q,h= - 1}(J/\psi)=0$. In Eq.~(\ref{vvp7hpm}), $\mu = m_b$, $C^{\mathrm{eff}}_{7\gamma} (\mu)=  C_{7\gamma} (\mu)- C_5(\mu)/ 3 - C_6(\mu)$,
$\alpha_e=1/129$ is the electromagnetic coupling constant and the scale $\nu$ refers to the $f_{J/\psi}$ decay constant scale. One also has 
$P_{3,5}^{q,h=\pm 1}(J/\psi)=0$ as well as $P_{3,5,7,9}^{q,(h=0)}(J/\psi)=0$.

\subsection{Vertex contributions}

In $B_s^0 \to \phi J/\psi$, the electroweak vertex receives $\alpha_s(\mu)$ corrections to all $a_{n}^{q,h}(\mu)$ in the amplitudes 
${\cal A}_{\B_s^0\to \phi J/\psi}^h$. For $h=0$, these are, with $\mu= m_b$, 
 \begin{equation}
  V^{h=0}_n(J/\psi) = \left\{
    \begin{array}{ll}
       \displaystyle
       12 \; {\rm ln} \left (\frac{m_b}{\mu}\right)  - 3 i \pi - \frac{27}{2}\ ,
      \medskip  \\ \qquad  \mathrm{for}  \quad n \in \{2,3,9\} &
       \\ [1em]
       \displaystyle
        -12 \; {\rm ln} \left (\frac{m_b}{\mu}\right)  + 3 i \pi + \frac{13}{2}\ ,
      \medskip  \\  \qquad  \mathrm{for}  \quad n \in \{5,7\}&
       \end{array}
  \right.
\end{equation}
 whereas for $h=-1$ one has,
 \begin{equation}
  V^{h=-1}_n(J/\psi) = \left\{
    \begin{array}{ll}
       \displaystyle
       12 \; {\rm ln} \left (\frac{m_b}{\mu} \right) + \pi^2 -\frac{143}{4}\ ,
       \medskip  \\ \qquad \mathrm{for} \,\, n \in \{2,3,9\} &
       \\  [1em]
       \displaystyle
       -12 \; {\rm ln}\left (\frac{m_b}{\mu} \right)- \pi^2+ \frac{95}{4}\ ,
       \medskip  \\  \qquad \mathrm{for} \quad n \in \{5,7\} &
     \end{array}
  \right.
\end{equation}
and for  $h=+1$ one has,
 \begin{equation}
  V^{h=+ 1}_n(J/\psi) = \left\{
    \begin{array}{ll}
       \displaystyle
       12 \; {\rm ln} \left (\frac{m_b}{\mu} \right) + \frac{\pi^2}{2} - 6 i \pi - \frac{71}{4}\ ,
       \medskip  \\ \qquad \mathrm{for} \,\, n \in \{2,3,9\} &
       \\ [1em]
       \displaystyle
       -12 \; {\rm ln}\left (\frac{m_b}{\mu} \right)- \frac{\pi^2}{2} + 6 i \pi + \frac{23}{4}\ ,
       \medskip  \\  \qquad \mathrm{for} \quad n \in \{5,7\}\ .  &
     \end{array}
  \right.
\end{equation}

\subsection{Hard scattering contributions}

The  gluon exchange between a $J/\psi$ meson  and the spectator $s$-quark  leads to the hard scattering amplitudes, 
\begin{equation}
   H_n^{h=0}(\phi J/\psi)=  \pm \ 3 \frac{B_{\phi J/\psi}}{A_{\phi J/\psi}^{h=0}}  \frac{m_{B_s^0}}{\lambda_{B_s^0}} \Big ( r_\chi^{\phi}(\mu) X_H + 3  \Big ) \! , 
  \end{equation}
for $h=0$, $\mu=m_b/2$ and $\lambda_{B_s^0}=0.350$ GeV~\cite{Beneke:2003zv}. The plus sign is for $n=2,3,9$ and the minus sign for $n=5,7$. 
The phenomenological amplitude, $X_H$, parametrizes  the endpoint divergence of the scalar meson's LCDA and is defined in Eq.~(\ref{defdiv}).  
For the helicity, $h=+ 1$, the correction reads 
\begin{equation}\hspace*{-2mm}
  H_n^{h=+1}(\phi J/\psi) =\mp \ 18 \frac{B_{\phi J/\psi}}{A_{\phi J/\psi}^{h=+1}} \frac{f_{\phi}^{\perp}}{f_{\phi}}  \frac{ m_{J/\psi}}{ \lambda_{B_s^0}} (X_H -1)\  ,      
\end{equation} 
where the minus sign applies to $n=2,3,9$ and the plus sign to $n=5,7$. The helicity, $h=-1$, contribution is simply,

\begin{equation}
  H_n^{h=-1}(\phi J/\psi) = 0 \;\; {\rm for} \;\; n=2,3,5,7,9\ .
\end{equation}

\begin{table*}[t]
\caption{\label{Wilsoncoefft}  Wilson coefficients at the $\mu=m_b$ and $\mu=m_b/2$ scales in the NDR scheme~\cite{Beneke:2001ev}. 
The coefficients $C_7(\mu)-C_{10}(\mu)$  must be multiplied by $\alpha_e$.}
\begin{center}
\begin{tabular}{|c|ccccccccccc|}
 \hline 
  & $C_1(\mu)$ & $C_2(\mu)$ & $C_3(\mu)$ & $C_4(\mu)$ & $C_5(\mu)$ & $C_6(\mu)$ & $C_7(\mu)$ & $C_8(\mu)$ & $C_9(\mu)$ & $C_{10}(\mu)$  &  $C_{7 \gamma}(\mu)$   \\ 
\hline 
\hline 
$\mu = m_b$ & $1.081$ & $-0.190$ & $0.014$ & $-0.036$ & $0.009$ & $-0.042$ &  $-0.011$&  $0.06$ & $-1.254$ & $0.233$ & $-0.318$ \\ 
$\mu = m_b/2$ & $1.137$ & $-0.295$ & $0.021$ & $-0.051$ & $0.010$ &  $-0.065$ &  $-0.24$ & $0.096$ & $-1.325$ & $0.331$ & $-0.364$  \\
\hline 
\end{tabular}
\end{center}
\end{table*}

\section{QCDF corrections for $\bm{B_s^0 \to f_0(980) J/\psi}$ decay amplitudes\label{secfour}}

We now turn to the $B_s^0 \to J/\psi f_0(980)$ transition for which the $\alphas(\mu)$ corrections are all included following Ref.~\cite{Beneke:2003zv}
applied to an $SV$ final state. For previously mentioned reasons, we solely employ the first non-vanishing leading term in the LCDA,
\begin{equation}
\phi_{f_0}(x)  =  6x(1-x) \Big [ 3 B_1 (\mu)(2x-1)  \Big ] \ , 
\end{equation}
where $B_1(m_b/2) =-0.54$~\cite{Cheng:2005nb} is the $f_0(980)$'s first Gegenbauer moment and we remind that only odd moments 
contribute in case of charge-neutral scalar mesons. In particular, contrary to the pseudoscalar LCDA, the leading term $6x(1-x) B_0$ vanishes since 
$B_0 = (m_1(\mu) - m_2(\mu))/m_S$, where $m_S$ is the scalar meson mass and $m_{1,2}(\mu)$ its running quark masses. The scalar
twist-3 two-particle distribution is given by
\be
 \varphi_{f_0}(x)  =1\ .
\en
The asymptotic forms of the LCDA, $\phi_{J/\psi}(x)$~ (Eq.~\eqref{lt2}) and $\varphi_{J/\psi}(x)$ (Eq.~\eqref{lt3}), are used.
As in the $B_s^0 \to \phi J/\psi$ decay, the $J/\psi$ meson is created from vacuum whereas the transition $B_s^0 \to f_0(980)$ produces the
scalar meson. Here, we only consider the $s\bar{s}$ component of the $f_0(980)$ since the flavor of the spectator quark in the tree and penguin 
topologies of $B_s^0$ decays is strange. There are no penguin corrections~\cite{Beneke:2003zv} to the $B_s^0 \to f_0(980) J/\psi$ decay 
amplitude in Eq.~\eqref{ASV}.

\subsection{Vertex contributions}
At the order of $\alpha_s(\mu)$, the vertex correction, $V_n(J/\psi)$, involves the leading twist distribution, $\phi_{J/\psi}(x)$, and a gluon kernel 
given in~\cite{Beneke:2003zv}.  We derive from this the expressions, 
 \begin{equation}
  V_n(J/\psi) = \left\{
    \begin{array}{ll}
       \displaystyle
       12 \; {\rm ln} \left (\frac{m_b}{\mu}\right)  - 3 i \pi - \frac{37}{2}\ ,
      \medskip \\ \qquad \mathrm{for} \,\, n \in \{2,3,9\} &
       \\[1em]
       \displaystyle
        -12\; {\rm ln} \left (\frac{m_b}{\mu}\right)  + 3 i \pi + \frac{13}{2}\ ,
      \medskip \\  \qquad \mathrm{for} \,\, n \in \{5,7\}  &
       \end{array}
  \right.
\end{equation}
with $\mu= m_b$.

\subsection{Hard scattering contributions}

The hard scattering correction in case of an $f_0 J/\psi$ final state reads
\begin{equation}\label{hsvs} 
  H_n(f_0 J/\psi) =   \pm\,  3\, \frac{B_{f_0 J/\psi}}{A_{f_0 J/\psi}} \frac{m_{B_s^0}}{\lambda_{B_s^0}} \Big ( \bar r_\chi^{f_0}(\mu) X_H + 3B_1(\mu)  \Big ) ,
  \end{equation}
where the plus sign applies to $n=2,3,9$, the minus sign to $n=5,7$ and $X_H$ is given, as in the case of the $\phi J/\psi$ final state, by Eq.~(\ref{defdiv}). 

The chiral coefficient, $\bar r_\chi^{f_0}(\mu)$, enters Eq.~(\ref{hsvs}) rather than $r_\chi^{f_0}(\mu)$ defined as,
\begin{equation}
\label{rchiSV}
r_\chi^{f_0}(\mu)= \frac{2 m_{f_0}^2}{m_b(\mu) \left(m_1(\mu) - m_2(\mu) \right)}\ .
\end{equation}
The reason is that in case of neutral scalar mesons, $m_1(\mu) = m_2(\mu)$ and $r_\chi^{f_0}(\mu)$ diverges. 
On the other hand, it is known from $C$-conjugation invariance that the vector decay constant of the neutral scalar meson must vanish. 
However, the quark equations of motions yield a relation between the scalar- and vector-decay constants, $\bar f_{f_0}$ and $f_{f_0}$ 
respectively: 
\begin{equation}
  \bar f_{f_0} = \frac{m_{f_0}}{m_1(\mu)-m_2(\mu)} f_{f_0} \ , 
\end{equation} 
where $m_{f_0} \bar f_{f_0} = \langle 0 | \bar q_2 q_1 | f_0  \rangle$. Since $\bar f_{f_0}$ is non-zero, the product $f_{f_0}m_{f_0} / (m_1(\mu) - m_2(\mu))$ 
is finite in the limit $m_1(\mu) \to m_2(\mu)$. We thus recombine, $f_{f_0}r_\chi^{f_0} = \bar  f_{f_0} \bar r_\chi^{f_0}$, with
\begin{equation}
 \bar{r}_\chi^{f_0}(\mu)= \frac{2 m_{f_0}}{m_b(\mu)}\ .
\end{equation}

\begin{table}[t*]
\begin{center}
\caption{Values of the higher order correction ($\rho_{C}$, $\phi_{C}$) and hard-scattering ($\rho_{H}$, $\phi_{H}$) parameters as function of the $B_s^0$ decay constant.}
\label{rhophi}
\renewcommand\arraystretch{1.6}
\begin{tabular}{|c|cccc|}
\hline
   $\fBs$ [MeV]     &   $\rho_{C}$   &   $\phi_{C}$ ($^\circ$)   &    $\rho_{H}$   &   $\phi_{H}$ ($^\circ$)     \\
\hline
\hline
  230                &  $ 4.52 \pm 2.24 $      & $ 173.8 \pm 37.6 $      &     $  1.90 \pm 0.20 $    &   $ 266.0 \pm 21.6 $           \\
  260                &  $ 6.16 \pm 2.03 $      & $ 176.1 \pm 53.6 $      &     $  1.70 \pm 0.16$     &   $ 260.6 \pm 19.3$            \\
  290                &  $ 7.33 \pm 1.63 $      & $ 176.0 \pm 57.6 $      &     $  1.54 \pm 0.15$     &   $ 255.6 \pm 17.3$           \\ 
 \hline  
\end{tabular} 
\end{center}
\end{table}

{\renewcommand\baselinestretch{1.6}
\begin{table*}[t]
\caption{Short-distance amplitudes, $a_n^{q,h}(m_b) \times 10^3$, for the helicity $h=+1$ in $B_s^0 \to J/\psi\phi$, as a function of the decay constant, $\fBs$, and with
$\bar{f}_{f_0^s}=380$ MeV. The $LOVP$ results are obtained with the leading order ($LO$) amplitude to which vertex $V$ and penguin $P$ corrections are added.
In case of $LOVPH$, the hard scattering contribution with the endpoint parametrization $X_H$ is included. $LOVPH+C$ contains additionally the purely phenomenological 
contribution $\zeta^{(h)}$ with two more parameters.}
\label{tabh=-1}
\begin{center}
\begin{tabular}{|c|c|cc|cc|cc|}
\hline 
     $\fBs$  [MeV]   &   &  \multicolumn{2}{|c|}{ 230}     &       \multicolumn{2}{|c|}{ 260}        &   \multicolumn{2}{|c|}{290}          \\ 
                              &  $LOVP$  &  $LOVPH$  & $LOVPH+C$     &   $LOVPH$  & $LOVPH+C$   &    $LOVPH$  &$LOVPH+C$   \\
 \hline \hline
 $a_2^{u,c}(m_b)$  & $60.38-i\,161.7$  & $-3.77+i\, 148.8$  & $-8.43 +i\,129.87$ & $38.06+i\, 149.7$ & $8.21 +i\,130.04$ & $75.52+i\, 140.4$ & $22.57 +i\,128.49$    \\
 $a_3^{u,c}(m_b)$  & $5.66+i\,5.39$ & $8.54-i\, 8.54$ & $ 8.75 -i\, 7.69$ & $6.66-i\, 8.58$ & $ 8.0 -i\, 7.70$ & $4.98-i\, 8.17$ & $7.36 -i\,7.63$    \\  
 $a_5^{u,c}(m_b)$ & $-5.27-i\,6.28$ & $-8.94+i\, 11.47$  & $-9.21 +i\,10.39$  & $-6.55+i\, 11.52$ & $-8.25 +i\,10.40$ & $-4.41+i\, 10.99$ & $-7.43 +i\,10.31$   \\
 $a_7^u(m_b)$  & $0.12+i\, 0.07$ & $0.17-i\, 0.13$ &  $0.17 -i\,0.12$ & $0.14-i\, 0.13$ & $0.16  -i\,0.12$ & $0.11-i\, 0.13$ & $0.15 -i\,0.12$     \\
 $a_7^c(m_b)$  & $0.69+i\, 0.07$ & $0.73-i\, 0.13$ &  $0.74 -i\,0.12$  & $0.71 -i\, 0.13$  & $0.73  -i\,0.12$  & $0.68-i\, 0.13$ & $0.72  -i\, 0.12$  \\                    
 $a_9^u(m_b)$  & $-9.25-i\, 0.27$ & $-9.40+i\, 0.43$ &  $-9.41 +i\,0.39$ & $-9.30+i\, 0.43$ & $-9.37 +i\,0.39$ & $-9.22+i\, 0.41$ & $-9.34 +i\,0.38$    \\
 $a_9^c(m_b)$  & $-8.68-i\, 0.27$&  $-8.83+i\, 0.43$  &  $-8.84 +i\,0.39$ & $-8.73+i\, 0.43$ & $-8.80 +i\,0.39$ & $-8.65+i\, 0.41$ & $-8.77 +i\,0.38$   \\
\hline  
\end{tabular}
\end{center}
\end{table*}}

{\renewcommand\baselinestretch{1.6}
\begin{table*}[t]
\caption{Short-distance amplitudes, $a_n^{q,h}(m_b) \times 10^3$, for helicity $h=-1$ and $B_s^0 \to J/\psi \phi$. Since the hard-scattering contributions are zero, these amplitudes are independent of $\fBs$.}
\label{tabh=+1}
\begin{center}
\begin{tabular}{|c|c|}
 \hline 
 $a_2^{u,c}(m_b)$   &  $-51.72$ \\
 $a_3^{u,c}(m_b)$   &   $9.39$ \\ 
 $a_5^{u,c}(m_b)$   &   $-9.63$ \\
 $a_7^{u,c}(m_b)$   &   $0.12$ \\  
 $a_9^{u,c}(m_b)$   &    $-9.49$ \\
\hline  
\end{tabular}
\end{center}

\end{table*}}

{\renewcommand\baselinestretch{1.6}
\begin{table*}[t]
\caption{As in Table~\ref{tabh=-1} but for the helicity $h=0$.}
\label{tabh=0}
\begin{center}
\begin{tabular}{|c|c|cc|cc|cc|}
\hline 
     $\fBs$  [MeV]   &  &    \multicolumn{2}{|c|}{ 230}     &       \multicolumn{2}{|c|}{ 260}        &   \multicolumn{2}{|c|}{290}            \\ 
                              &   $LOVP$  &   $LOVPH$  & $LOVPH+C$     &   $LOVPH$  & $LOVPH+C$   &    $LOVPH$  &$LOVPH+C$   \\
 \hline \hline
 $a_2^{u,c}(m_b)$   & $54.51-i\, 80.86$ & $160.2-i\, 132.6$ & $161.0 -i\, 129.4$  & $165.7-i\,132.7$ & $170.6 -i\,129.5$ & $171.8-i\, 131.2$ & $180.6 -i\,129.2$  \\
 $a_3^{u,c}(m_b)$   & $5.86+i\, 2.69$ & $1.11+i\, 5.01$ &  $1.08 +i\,4.87$ & $0.87+i\, 5.02$ &   $0.65 +i\,4.87$ & $0.60+i\, 4.95$ & $0.20 +i\,4.86$  \\
 $a_5^{u,c}(m_b)$   & $-7.17-i\, 3.14$ & $-1.12-i\, 6.10$  & $-1.08  -i\,5.92$ & $-0.81-i\, 6.11$ &  $-0.53 -i\,5.92$ & $-0.46-i\, 6.02$ & $0.04 -i\,5.90$  \\ 
 $a_7^{u,c}(m_b)$   & $0.09+i\, 0.03$ & $0.02+i\, 0.07$ & $0.02 +i\,0.07$  & $0.02+i\, 0.07$   &  $0.02 +i\,0.07$ & $0.02+i\, 0.07$ & $0.01 +i\,0.06$  \\ 
 $a_9^{u,c}(m_b)$   & $-9.31-i\,0.14$ & $-9.07-i\, 0.25$ & $-9.07 -i\,0.24$  & $-9.06-i\, 0.25$   &  $-9.05 -i\,0.24$ & $-9.05-i\, 0.25$ &  $-9.03 -i\,0.24$  \\
\hline  
\end{tabular}
\end{center}
\end{table*}}

\section{Numerical parameters \label{secfive}}

This section serves to summarize all parameter values required for numerical applications. The Wilson coefficients at the scales $\mu=m_b$ and $\mu=m_b/2$
used in this work are listed in Table~\ref{Wilsoncoefft}. For the meson masses, we refer to the latest PDG values~\cite{Amsler:2008zzb}, which are (in GeV):
\begin{multline}
m_{B_s^0}= 5.366\ , \; \; m_{B_s^{\star}}= 5.412\ , \; \; m_{f_0}=0.980 \ ,\\    
m_{J/\psi}= 3.096\ ,  \;\; m_{\phi}= 1.019\ . \hspace*{1cm}
\end{multline}
The running quark masses at $\mu = m_b =4.2$ GeV are (in GeV),
 \be
m_{b}= 4.2\, , \; m_{c}= 1.3\, , \; m_{s}= 0.07\, , \; m_{u,d}=0.003\, ,
\en
and those at $\mu = m_b/2 =2.1$~GeV are,
\be
\hspace*{-1mm}
m_{b}= 4.95\, , \; m_{c}= 1.51\, , \; m_{s}= 0.09\, , \; m_{u,d}=0.005\ .
\en 

We take the $\phi$ decay constant values from Ref.~\cite{Beneke:2006hg}: $f_{\phi}= (221 \pm 3)$~MeV and $f_{\phi}^{\perp}= (175 \pm 25)$~MeV.  
For the $J/\psi$ meson, we use $f_{J/\psi}= (416 \pm 6)$~MeV~\cite{Hwang:2006cua} and $f_{J/\psi}^{\perp}=(405 \pm 5)$~MeV~\cite{Cheng:2001ez}.
In the $B_s \to J/\psi f_0(980)$ channel, the $s\bar{s}$ component of the $f_0(980)$ is involved which implies the poorly known scalar decay constant 
$\bar f_{f_0}$: one theoretical estimate yields $\bar f_{f_0^s} = (180 \pm 15)$~MeV~\cite{DeFazio:2001uc} whereas a much larger value 
$\bar f_{f_0^s} (1\, \mathrm{GeV}) = (370\pm 20)$~MeV $\big [ \bar f_{f_0^s} (2.1~\mathrm{GeV}) = (460 \pm 25)\mathrm{MeV} \big ]$ is found in  
Ref.~\cite{Cheng:2005nb}, both from coupling to the scalar $\bar ss$ current  only (denoted by the superscript $s$ in $f_0^s$, which we use henceforth). 
Similarly, several theoretical predictions exist for the leptonic $B_s$ decay constants of which we select three values from unquenched lattice QCD:
 $\fBs = (204 \pm 12^{+24}_{-23})$~MeV~\cite{Lellouch:2000tw}, $\fBs = (259 \pm 32)$~MeV~\cite{Gray:2005ad} and $\fBs = (231 \pm 15)$~MeV~\cite{Gamiz:2009ku}. 

To illustrate the sensitivity of the ratio $\mathcal{R}_{f_0/\phi}$ to the hadronic uncertainties, we exemplarily choose three different values for each decay 
constant:  $\fBs = 230$, $260$, $290$~MeV and $\bar f_{f_0^s}=340$, $380$, $420$~MeV.

\subsection{$\bm{B \to V}$ transition form factor \label{bvform}}

Values for the $B_s^0 \to \phi$ transition form factors are taken from the pole-extrapolation model by Melikhov~\cite{Melikhov:2001zv}:
 \be\label{ffa0v}
A_0(q^2)^{B_s^0 \to \phi} = \frac{a_0(0)}{\left(1-\frac{q^2}{m^2_{B_s^0}}\right) \left(1-\sigma_1 \frac{q^2}{m^2_{B_s^0}}
   + \sigma_2 \frac{q^4}{m^4_{B_s^0}}   \right)}\ .
\en
The form factor $V(q^2)^{B_s^0 \to \phi}$ is given by a similar expression in which $a_0(0)$ is replaced by $v(0)$ and $m_{B_s^0}$ 
by $m_{B_s^{\star}}$~\cite{Melikhov:2001zv}.  Next, the $A_1(q^2)^{B_s^0 \to \phi}$ form factor is parametrized by
\be\label{ffa1a2}
A_1(q^2)^{B_s^0 \to \phi}=\frac{a_1(0)}{\left(1-\sigma_1 \frac{q^2}{m^2_{B_s^{\star}}}+\sigma_2 \frac{q^4}{m^4_{B_s^{\star}}}    \right)}\ .
\en
Finally, $A_2(q^2)^{B_s^0 \to \phi}$ has the same functional form as  $A_1(q^2)^{B_s^0 \to \phi}$ where $a_1(0)$ is replaced 
by  $a_2(0)$. In both, Eqs. \eqref{ffa0v} and \eqref{ffa1a2}, the momentum transfer is $q^2 = m^2_{J/\psi}$. 

In Eqs.~\eqref{ffa0v} and \eqref{ffa1a2}, the form factors at $q^2=0$ are $a_0(0)=0.42$ ($v(0)=0.44$) and $a_1(0)=0.34$ ($a_2(0)=0.31$). 
The extrapolation parameters are, for $A_0(q^2)^{B_s^0 \to \phi}$,  $\sigma_1=0.55$ and $\sigma_2=0.12$; for  $V(q^2)^{B_s^0 \to \phi}$, 
$\sigma_1=0.62$ and $\sigma_2=0.20$; for $A_1(q^2)^{B_s^0 \to \phi}$, $\sigma_1=0.73$ and $\sigma_2=0.42$  and finally for  
$A_2(q^2)^{B_s^0 \to \phi}$, $\sigma_1=1.30$ and $\sigma_2=0.52$. The respective values for the form factors at the value 
$q^2=m_{J/\psi}^2$ are $A_0(q^2)^{B_s^0 \to \phi}=0.76$, $A_1(q^2)^{B_s^0 \to \phi}=0.42$, $A_2(q^2)^{B_s^0 \to \phi}=0.49$ and 
$V(q^2)^{B_s^0 \to \phi}=0.80$.

\subsection{$\bm{B \to S}$ transition form factor \label{bsform}}

We studied the transition form factor, $F_{0,1}^{B_s^0 \to f_0^s}(q^2)$, in a comparative calculation using a dispersion relation and a covariant light front dynamics 
model~\cite{ElBennich:2008xy}. To our knowledge, this form factor has only been calculated recently in QCD sum rules~\cite{Ghahramany:2009zz,Colangelo:2010bg} 
and pQCD~\cite{Li:2008tk} for $q^2=0$ and must be extrapolated to the value $F_{0,1}^{B_s^0 \to f_0^s}(m_{J/\psi}^2)$.

In our work~\cite{ElBennich:2008xy}, the transition form factors are derived from the constituent quark three-point function, the vertices of which are the 
weak interaction coupling, $\gamma_\mu(1-\gamma_5)$, and two phenomenological Bethe-Salpeter amplitudes for the $B_{(s)}$ and $f_0(980)$ mesons. 
While the $B_s$ can be parametrized with the leptonic decay constant (known from lattice-QCD simulations), the latter is more problematic since the  
$\bar f_{f_0^s}$ is poorly determined. In an attempt to formulate a suitable scalar $f_0(980)$ vertex function, we constrained its parameters by means of 
experimental quasi two-body branching fractions, $D_{(s)}\to f_0 (980) P$, $P=\pi, K$. The advantage is that the $F_+^{B_s^0 \to f_0^s}(q^2)$ and 
$F_-^{B_s^0 \to f_0^s}(q^2)$ form factors,
\begin{eqnarray}
 \lefteqn{\hspace*{-8mm} \langle f_0^s (p_2) | \bar s \gamma_\mu(1-\gamma_5) b | B_s^0 (p_1)\rangle\  = }  \nonumber \\
  & &\hspace*{-1cm}   F_+^{B_s^0 \to f_0^s}(q^2) (p_1+p_2)_\mu + F_-^{B_s^0 \to f_0^s}(q^2) (p_1-p_2)_\mu ,
\end{eqnarray}
can be calculated for any physical time-like momentum transfer $q^2=(p_1-p_2)^2$. The superscript $s$ is a reminder that the transition is to the $\bar ss$ 
component of the scalar meson and $p_1$ and $p_2$ are the $B_s^0$ and $f_0(980)$ four-momenta, respectively. We do stress that the $B_s \to f_0(980)$ 
form factor calculated by us in Ref.~\cite{ElBennich:2008xy} does not assume a pure $\bar ss$ state of the $f_0(980)$. Instead, it was treated as a mixture of
strange and non-strange $\bar qq$ components related by a mixing angle which also yields the related form factor $F_{0,1}^{B \to f_0^{u,d}}(q^2)$.  
This angle was determined with experimental constraints \cite{ElBennich:2008xy} and the overall normalization of the transition form factor receives 
contributions from both states. 

 The form factors $F_\pm(q^2)$
(we suppress the flavor superscripts) are related to the set of vector and scalar form factors as, 
\begin{eqnarray}
  F_1(q^2) & = & F_+(q^2)\ ,  \\ 
  F_0(q^2) & = & F_+(q^2) + \frac{q^2}{m_{B_s^0}^2 - m_{f_0}^2} F_-(q^2)\ .  
\end{eqnarray}
The form factor $F_1(q^2)$ we obtain in both the dispersion relation and  covariant light front dynamics approaches agree at the maximum recoil 
point $q^2=0$. At large
four-momentum transfer, specifically for $q^2=m_{J/\psi}^2\simeq 10$~GeV$^2$, our model predictions differ significantly which is also known to 
occur for $B\to \pi$ transition form factors~\cite{ElBennich:2009vx}. This is not surprising, as for large momentum transfers the final-state meson 
is less energetic and the soft physics of the bound states becomes more relevant. Since the models differ in their parametrization of the bound-state 
wave functions, it is clear that their inaccuracies are revealed in the form-factor predictions at large $q^2$. In Ref.~\cite{Colangelo:2010bg}, we
deduce from the author's extrapolation parametrization that $F_1^{B_s^0 \to f_0^s}(m_{J/\psi}^2)\simeq 0.3$, which is compatible with our 
dispersion-relation prediction $\simeq 0.4$ within the errors. In Section \ref{secsix}, we will account for this rather large window of values and 
plot the ratio $\mathcal{R}_{f_0/\phi}$ as a function of  $F_1^{B_s^0 \to f_0^s}(m_{J/\psi}^2)$.

\subsection{Model parameters \label{modelparam}}

The hard scattering contributions involve endpoint divergences, which we choose to parametrize by,
\begin{equation} \label{defdiv}
X_H = \Bigl ( 1+ \rho_H \exp (i \phi_H) \Bigr)  \ln \frac{m_{B_s^0}}{\lambda_h}\ .
\end{equation}
In case of a possible annihilation or ``other"  contribution we simply write, 
\begin{equation}\label{defdivann}
X_C =  \rho_C \exp(i \phi_C) 
\end{equation}
which introduces four parameters, $0 < \rho_{C,H}$ and $0 < \phi_{C,H} < 360^{\circ}$. We assume that $X_{C,H}^{h=0}=X_{C,H}^{h=-1}=X_{C,H}^{h=+1}=X_{C,H}$, 
as the vector $\phi$ and scalar $f_0(980)$ mesons have similar masses and we consider the $s\bar s$ component only. The  hard scattering corrections
are expected to be of the order of $m_{B_s^0}/\lambda_h$ in Eq.~(\ref{defdiv}), with $\lambda_h=0.5$~GeV. The parameters $\rho_{C,H}$ and 
$\phi_{C,H}$ are chosen so as to reproduce the experimental data discussed in Section~\ref{secsix}.  We insert their values in the $B_s^0 \to J/\psi f_0$ 
decay amplitude~\eqref{ASV} and then predict the branching ratio $\mathcal{B}(B_s^0 \to f_0 J/\psi  )$.  

{\renewcommand\baselinestretch{1.6}
\begin{table*}[t]
\caption{Short-distance amplitudes, $a_n^q(m_b) \times 10^3$, for  $B_s^0 \to  J/\psi f_0(980)$ as a function of the  $\fBs$ 
decay constant  with $\bar{f}_{f_0^s}=380$~MeV and $F_1^{B_s^0 \to f_0^s}(m_{J/\psi}^2)=0.4$.
See caption in Table~\ref{tabh=-1} for the definition of  $LOVP$, $LOVPH$ and $LOVPH+C$ amplitudes.}
\label{tabf0}
\begin{center}
\begin{tabular}{|c|c|cc|cc|cc|}
\hline 
     $\fBs$  [MeV]   &  &    \multicolumn{2}{|c|}{ 230}     &       \multicolumn{2}{|c|}{ 260}        &   \multicolumn{2}{|c|}{290}            \\ 
                              &   $LOVP$  &   $LOVPH$  & $LOVPH+C$     &   $LOVPH$  & $LOVPH+C$   &    $LOVPH$  &$LOVPH+C$   \\
 \hline \hline
 $a_2^{u,c}(m_b)$   & $11.61-i\, 80.86$ & $-42.40-i\, 255.5$  & $-33.35  -i\,224.3$ & $-66.51-i\, 234.1$ &  $-51.82 -i\,224.4$ & $-95.23-i\, 229.5$ & $-69.17  -i\,223.8$  \\
 $a_3^{u,c}(m_b)$   & $7.29+i\,2.69$ & $9.71+i\, 10.53$  & $9.30 +i\,9.13$ & $10.80+i\, 9.57$ & $10.13 +i\,9.13$ & $12.08+i\, 9.36$ & $10.91 +i\,9.10$  \\   
 $a_5^{u,c}(m_b)$   & $-7.17-i\, 3.14$ & $-10.25-i\, 13.12$ & $-9.74  -i\,11.34$  & $-11.63-i\, 11.90$ & $-10.79 -i\,11.35$ & $-13.27-i\, 11.64$ & $-11.78 -i\,11.31$  \\  
 $a_7^{u,c}(m_b)$   & $0.09+0.03$ & $0.13+i\, 0.15$  & $0.12 +i\,0.13$  & $0.14+i\, 0.14$  &  $0.14 +i\,0.13$ & $0.16+i\, 0.13$ & $0.15 +i\,0.13$  \\     
 $a_9^{u,c}(m_b)$   & $-9.38-i\, 0.14$ & $-9.51 -i\, 0.53$ & $ -9.49  -i\,0.46$  & $-9.56-i\, 0.48$ &  $-9.53  -i\,0.46$ & $-9.63-i\, 0.47$ & $-9.57 -i\,0.46$  \\ 
\hline  
\end{tabular}
\end{center}
\end{table*}}

{\renewcommand\baselinestretch{1.6}
\begin{table*}[t]
\caption{The phenomenological contributions  $\zeta^h \times 10^3$ for $h=0,-1,+1$, Eq.~\eqref{bdef}, to the $B_s^0 \to J/\psi \phi$ amplitude as a function of the $\fBs$  decay constant with $\bar{f}_{f_0^s}=380$ MeV.}
\label{tabannphi}
\begin{center}
\begin{tabular}{|c|ccc|}
\hline 
     $\fBs$ [MeV]      &         230                      &          260                              & 290                 \\
\hline\hline 
 $\zeta^{h=0}$            &  $-18.11 +i\,1.98$  &  $-28.04 +i\,1.89$ &  $-37.19 +i\,2.63$ \\

\hline 

 $\zeta^{h=-1}$            &  $-129.26 +i\,14.12$  &  $-200.12 +i\,13.46$ &  $-265.41 +i\,18.77$ \\  
\hline 

 $\zeta^{h=+1}$            &  $-15.25 + i\,1.67$  &  $-23.61 + i\,1.59$ &  $-31.31 +i\,2.21$ \\  
\hline  
\end{tabular}
\end{center}

\end{table*}}

{\renewcommand\baselinestretch{1.6}
\begin{table*}[t]
\caption{Same as Table~\ref{tabannphi} but for the $B_s^0 \to  J/\psi f_0(980)$ amplitude.}
\label{tabannf0}
\begin{center}
\begin{tabular}{|c|ccc|}
\hline 
     $\fBs$ [MeV]      &         230                      &          260                              & 290                 \\
 \hline 
\hline
 $\zeta$            &  $-44.08 +i\,4.81$  &  $-68.25 +i\,4.59$  &  $-90.51 +i\,6.40$  \\ 
\hline 
\end{tabular}
\end{center}
\end{table*}}

{\renewcommand\baselinestretch{1.6}
\begin{table*}[t]
\caption{Prediction for the $B_s^0 \to J/\psi f_0$ observables for the different amplitudes $LOVP$, $LOVPH$ and $LOVPH+C$ along with experimental analysis data  
of the  $B_s^0 \to J/\psi \phi$ decay. Here central values, $\fBs=260$~MeV and $\bar f_{f_0^s} =380$~MeV, and the transition 
form factor $F_1^{B_s^0 \to f_0^s}(q^2=m^2_{J/\psi}) = 0.4$ are used. The values in the second column are predictions. Those of the third column include the hard
scattering corrections with the endpoint parametrization $\rho_H=1.85\pm0.07$ and $\phi_H=255.9^\circ\pm24.6^\circ$. The fourth column corresponds to the 
reproduction of the data with the parameters $\rho_H$, $\phi_H$, $\rho_C$ and $\phi_C$ displayed in the second line of Table~\ref{rhophi}.} 
\label{sumfit}
\begin{center}
\begin{tabular}{|c|c|c|c|c|}
\hline 
      &   $LOVP$    &    $LOVPH$  &       $LOVPH+C$        &    Experimental        \\ 
      &   (Prediction)         &   (2 parameters)         &    (4 parameters)  &   data   \\
 \hline \hline
 $\abs{\Amp_L}^2$   &$0.172$ & $0.554$ & $0.555$ & $0.555\pm 0.033$~\cite{Abazov:2008jz}  \\
$\abs{\Amp_\parallel}^2$ & $ 0.404$ & $0.219$ & $0.244$ &  $0.244 \pm 0.046$~\cite{Abazov:2008jz} \\
 $\phi_{\parallel}(\mathrm{rad})$ & $-0.221$ & $2.13$ & $2.72$ & $2.72\pm1.38$~\cite{Abazov:2008jz}  \\
 $\mathcal{B}(B_s^0 \to  J/\psi \phi )$& $ 0.00075$ & $0.00115$ & $0.00093$ & $0.00093\pm0.00033$~\cite{Amsler:2008zzb}\\
  $\mathcal{B}(B_s^0 \to J/\psi f_0)$ &$ 0.00020$ &  $0.00047$ & $0.00050$ &\\
  $A_{CP}(B_s^0 \to J/\psi f_0)$ & $-0.00013$ &$ -0.0013$  & $-0.0011$&\\
  $\mathcal{R}_{f_0/\phi}$ & $0.28$ & 0.42& $0.55$& \\
\hline  
\end{tabular}
\end{center}
\end{table*}}

\section{Results and experimental data \label{secsix}}

In the $B_s^0 \to \phi J/\psi$ decay, one can define five observables: a longitudinal, parallel and perpendicular polarization fraction, $f_L, f_{\parallel}$
and $f_{\perp}$, respectively, 
\begin{equation}
  f_{k} = \frac{\abs{\Amp_{k}}^2}{
  \abs{\Amp_L}^2+\abs{\Amp_\parallel}^2+\abs{\Amp_\perp}^2}\ , \; k=L,\parallel,\perp
\end{equation}
as well as two relative phases, $\phi_{\parallel}$ and $\phi_{\perp}$, 
\begin{equation}
  \phi_{k} = \arg \left( \frac{\Amp_{k}}{\Amp_L}\right)\ , \; k=\parallel,\perp \ ,
\end{equation}
where we have abbreviated,
$\Amp_{L}={\cal A}_{\B_s^0\to \phi J/\psi}^{(h=0)}$, $\Amp_{\parallel} = \big [ {\cal A}_{\B_s^0\to \phi J/\psi}^{(h=+1)}+
 {\cal A}_{\B_s^0\to \phi J/\psi}^{(h=-1)} \big ] /\sqrt{2}$ 
and  $\Amp_{\perp}= \big [{\cal A}_{\B_s^0\to \phi J/\psi}^{(h=+1)}-{\cal A}_{\B_s^0\to \phi J/\psi}^{(h=-1)} \big ]/\sqrt{2}$.  

\vspace*{1mm}
The $CP$ average is defined in terms of the polarization fractions, $f_k$, 
\begin{equation}
    A_{\CP}^{k} = \frac{f_k^{\Bbar_s^0}-f_k^{\B_s^0}}{f_k^{\Bbar_s^0}+f_k^{\B_s^0}}\ .
\end{equation}
Similarly, for $B_s^0 \to f_0(980) J/\psi$, the $CP$ average is defined as,
\begin{equation}\label{acpf0}
    A_{\CP} = \frac{\mathcal{B}(\Bbar_s^0 \to f_0 J/\psi) -\mathcal{B}(\B_s^0 \to f_0 J/\psi)}{\mathcal{B}(\Bbar_s^0 \to f_0 J/\psi) +\mathcal{B}(\B_s^0 \to f_0 J/\psi)}\ .
\end{equation}

We use data from CDF and D$\emptyset$ for the $B_s^0 \to \phi J/\psi$ decay, whereas there is no available data on the channel $B_s^0 \to  f_0 J/\psi $. 
Our  data compilation consists of the D$\emptyset$ values for the amplitudes, $\abs{\Amp_L}^2=0.555 \pm 0.027 \pm 0.006$, 
$\abs{\Amp_\parallel}^2=0.244 \pm 0.032 \pm 0.014$ and the relative phase  $\phi_{\parallel}=2.72^{+1.12}_{-0.27}$~rad~\cite{Abazov:2008jz}. 
The CDF values~\cite{Kuhr:2007dt} are compatible, $\abs{\Amp_L}^2=0.530 \pm 0.021 \pm 0.007$ and $\abs{\Amp_\parallel}^2=0.230 \pm 0.027 \pm 0.009$, 
and the PDG data book quotes the branching fraction, $\mathcal{B}(B_s^0 \to  J/\psi  \phi)=(9.3 \pm 3.3) \times 10^{-4}$~\cite{Amsler:2008zzb}.  

The ratio $\mathcal{R}_{f_0/\phi}$ has been argued~\cite{Stone:2008ak} to be of the order $0.2-0.3$, based on the knowledge of the experimental ratio 
of decay rates~\cite{Frabetti:1995sg},
\begin{equation} \hspace*{-1.8mm}
 \frac{\Gamma (D_s^+ \! \to \! f_0 \pi^+ \! \to K^+ K^- \pi^-)}{\Gamma (D_s^+ \! \to \phi \pi^+ \!\! \to K^+ K^- \pi^-)} = 0.28\pm 0.12 ,
\end{equation}
and an estimate of the semileptonic, integrated branching fraction ratio 
$\mathcal{B}(D_s^+ \to f_0 e^+ \nu, f_0 \to \pi^+\pi^- ) / \mathcal{B}(D_s^+ \to \phi e^+ \nu, \phi \to K^+K^-)  = (13\pm 4)\%$ from CLEO~\cite{Yelton:2009cm}. 
The ratio $\mathcal{R}_{f_0/\phi}$ was reassessed in terms of the differential decay ratio~\cite{Ecklund:2009fia},
\begin{eqnarray}
\label{rf03}
  \mathcal{R}_{f_0/\phi} & = & \frac{ \frac{d\Gamma}{dq^2} (D_s^+ \to f_0 e^+ \nu, f_0 \to \pi^+\pi^- )\big |_{q^2 = 0} }
  { \frac{d\Gamma}{dq^2} (D_s^+ \to \phi e^+ \nu, \phi \to K^+K^-)\big |_{q^2 = 0}  } \nonumber \\     & =  & 0.42 \pm 0.11.
\end{eqnarray}
If we combine the above three experimental estimates, we propose a window of $0.2 \lesssim  \mathcal{R}_{f_0/\phi} \lesssim  0.5$ for the ratio based on $D_s$ decays.

With the experimental data listed under Eq.~\eqref{acpf0} as constraint, we find optimal values for $X_C$ and $X_H$. In principle, we deal with a system
of four coupled non-linear equations for $\abs{\Amp_L}^2$, $\abs{\Amp_\parallel}^2$, $\phi_{\parallel}$ and $\mathcal{B}(B_s^0 \to  \phi J/\psi)$ and four 
variables, which does not put tight constraints on the phenomenological part of our $B_s^0 \to J/\psi \phi$ amplitude. When solving numerically we find, 
depending on the $\fBs$ values, two solutions among which only one yields a reasonable value for the branching fraction 
$\mathcal{B} (B_s \to f_0 J/\psi)$ not too different from that in a naive quark model. We list the parameters $\rho_{C,H}$ and $\phi_{C,H}$ 
independent of $\bar f_{f_0^s}$  for three values of $\fBs$ in Table~\ref{rhophi}, from which it is plain that the uncertainties on the magnitude of the modulus 
$\rho_C$  as well as the phase $\phi_C$ are substantial. The experimental errors on the observables are clearly not constraining enough. Yet, we observe 
that the variations of $X_C$  and $X_H$ are smooth as a function of the decay constant $\fBs$.

Likewise, we present numerical values for $a_n^{q,h}(m_b)$ for the three helicities in $B_s^0 \to J/\psi \phi $ in Tables~\ref{tabh=-1}, \ref{tabh=+1} and 
\ref{tabh=0} and for $B_s^0 \to J/\psi f_0$ in Table~\ref{tabf0} as functions of $\fBs$ to illustrate one facet of the hadronic uncertainty. In these tables, we list 
the decomposition of $a_n^{q,h}(m_b)$ for each value of $\fBs$; in the first column, the values of $a_n^{q,h}(m_b)$ are for the calculated leading order 
($LO$), vertex ($V$) and penguin ($P$) amplitudes only. These are independent of $\fBs$ and correspond to the predictions in Figure~\ref{fig1}. 
Next, the $a_n^{q,h}(m_b)$ that contain the $LO$,  $V$, $P$ and the hard-scattering ($H$) amplitudes, where only $\rho_H$ and $\phi_H$ are fitted to 
reproduce the $B_s^0 \to  \phi J/\psi$ observables while $X_C=0$.  For $\fBs=260$~MeV one obtains $\rho_H=1.85\pm0.07$ and $\phi_H=255.9\pm24.6^¡$. 
These values are not very different from those given in the second line of Table~\ref{rhophi}. This case corresponds to Figure~\ref{fig2}. At last, denoted 
by $LOVPH+C$, we give the values for $a_n^{q,h}(m_b)$ for the case  that the  $\zeta^{(h)}$ amplitudes are included, which corresponds to the $\rho_{C,H}$ 
and $\phi_{C,H}$ values in Table~\ref{rhophi} and to Figure~\ref{fig3}. We remind that the dependence on $\fBs$ enters the short-distance coefficients via the 
hard-scattering contribution $H_n^h(M_1 J/\psi)$ in Eq.~\eqref{defa} and that the phenomenological amplitudes, $X_H$ and $X_C$, are in competition 
with each other. Therefore, the hard scattering contributions to $a_n^{q,h}(m_b)$ in $LOVPH$ are slightly different than those to $LOVPH+C$.

The largest values observed in the leading amplitude, $a_2^{u,c}(m_b)$, are for $h=0$. We also remark there is no variation as a function of $f_{B_s}$ 
in Table~\ref{tabh=+1} since $H_n^{h=-1}(M_1 J/\psi)=0$.  Moreover, penguin contractions only contribute to $a_7^{q, h=+1}(m_b)$ and $a_9^{q, h=+1}(m_b)$ 
in the $B_s^0 \to  \phi J/\psi $ amplitudes, while there are no penguin terms in $B_s^0 \to f_0 J/\psi $. Altogether, the penguin contributions are very small.
We note that the contribution of the phenomenological amplitudes, $\zeta^{(h)}$ (Tables~\ref{tabannphi} and \ref{tabannf0}), is small, about $6-7\%$ of the 
$h=0,+1$ amplitudes in $B_s \to \phi J/\psi$  and $2\%$ of the $B_s \to f_0 J/\psi$ amplitude, yet dominant in the $h=-1$ amplitude devoid of penguin and 
hard scattering corrections. Thus, any contribution from new physics, and to less an extent annihilation topologies, should occur in the $h=-1$ helicity amplitude.

When including all the contributions ($LOVPH+C$), we qualitatively verify the hierarchy relation, 
$|\mathcal{A}^{(h=0)}_{B_s^0 \to \phi J/\psi} | > | \mathcal{A}^{(h=+1)}_{B_s^0 \to \phi J/\psi} | > | \mathcal{A}^{(h=-1)}_{B_s^0 \to \phi J/\psi} | $, in $B_s^0\to J/\psi\phi$
and $| \mathcal{A}^{(h=0)}_{B_s^0 \to \phi J/\psi} | > |  \mathcal{A}^{(h=-1)}_{B_s^0 \to \phi J/\psi} | > | \mathcal{A}^{(h=+1)}_{B_s^0 \to \phi J/\psi} | $ in the 
$CP$ conjugate decay $\bar B_s^0 \to J/\psi \phi$. These hierarchy relations are also reproduced for the amplitudes when they include, besides tree 
contributions, vertex, penguin and hard-scattering corrections.

Having determined numerical values for $X_H$ and $X_C$, we can calculate the $B_s^0 \to f_0 J/\psi$ amplitude and obtain the associated branching fraction
and $CP$ asymmetry.
 We do so for the central values of $\fBs=260$~MeV and $\bar f_{f_0^s} =380$~MeV discussed in Section~\ref{secfive}. 
For a transition form factor $F_1^{B_s^0 \to f_0^s}(q^2=m^2_{J/\psi}) = 0.4$ and for the different amplitudes $LOVP$, $LOVPH$ and $LOVPH+C$ defined above, those observables are displayed in Table~\ref{sumfit} together with a comparison of  the $B_s^0 \to J/\psi \phi $ results with the corresponding available experimental analysis values.
Furthermore, we obtain
for a transition form factor $F_1^{B_s^0 \to f_0^s}(q^2=m^2_{J/\psi}) = 0.2$:
\begin{eqnarray*}
\mathcal{B} (B_s \to f_0 J/\psi)  & =  & 3.80\times 10^{-4} , \\
A_{\mathrm{CP}} (B_s \to f_0 J/\psi)  & = & -0.0005\ ,  \\
\mathcal{B} (B_s \to \phi  J/\psi)  &  =  & 9.30\times 10^{-4}  , \\
\mathcal{R}_{f_0/\phi} & = & 0.42 \ ;
\end{eqnarray*}
for $F_1^{B_s^0 \to f_0^s}(q^2=m^2_{J/\psi}) = 0.3$,
\begin{eqnarray*}
\mathcal{B} (B_s \to f_0 J/\psi)  & =  &  4.37\times 10^{-4} , \\
A_{\mathrm{CP}} (B_s \to f_0 J/\psi)  & = & -0.0008\ ,  \\
\mathcal{B} (B_s \to \phi  J/\psi)  &  =  & 9.30\times 10^{-4}  , \\
\mathcal{R}_{f_0/\phi} & = & 0.48  \ ;
\end{eqnarray*}
and for $F_1^{B_s^0 \to f_0^s}(q^2=m^2_{J/\psi}) = 0.5$,
\begin{eqnarray*}
\mathcal{B} (B_s \to f_0 J/\psi)  & =  &  5.7\times 10^{-4} , \\
A_{\mathrm{CP}} (B_s \to f_0 J/\psi)  & = & -0.0013\ ,  \\
\mathcal{B} (B_s \to \phi  J/\psi)  &  =  & 9.30\times 10^{-4}  , \\
\mathcal{R}_{f_0/\phi} & = & 0.63  \ ,
\end{eqnarray*} 
and finally, the $CP$ asymmetries in $B_s \to J/\psi \phi$ are,
\begin{eqnarray*}
A_{\mathrm{CP}}^{L} (B_s \to \phi J/\psi)       & = &  -1.66 \times 10^{-3}\ ,  \\
A_{\mathrm{CP}}^{\parallel} (B_s \to \phi  J/\psi)     & = & 1.99\times 10^{-3}\ ,  \\
A_{\mathrm{CP}}^{\perp} (B_s \to \phi  J/\psi)  & = & 2.15 \times 10^{-3}\ .  \\
\end{eqnarray*}

{\renewcommand\baselinestretch{1.1}
\begin{figure*}[t!]
\includegraphics*[scale=0.54]{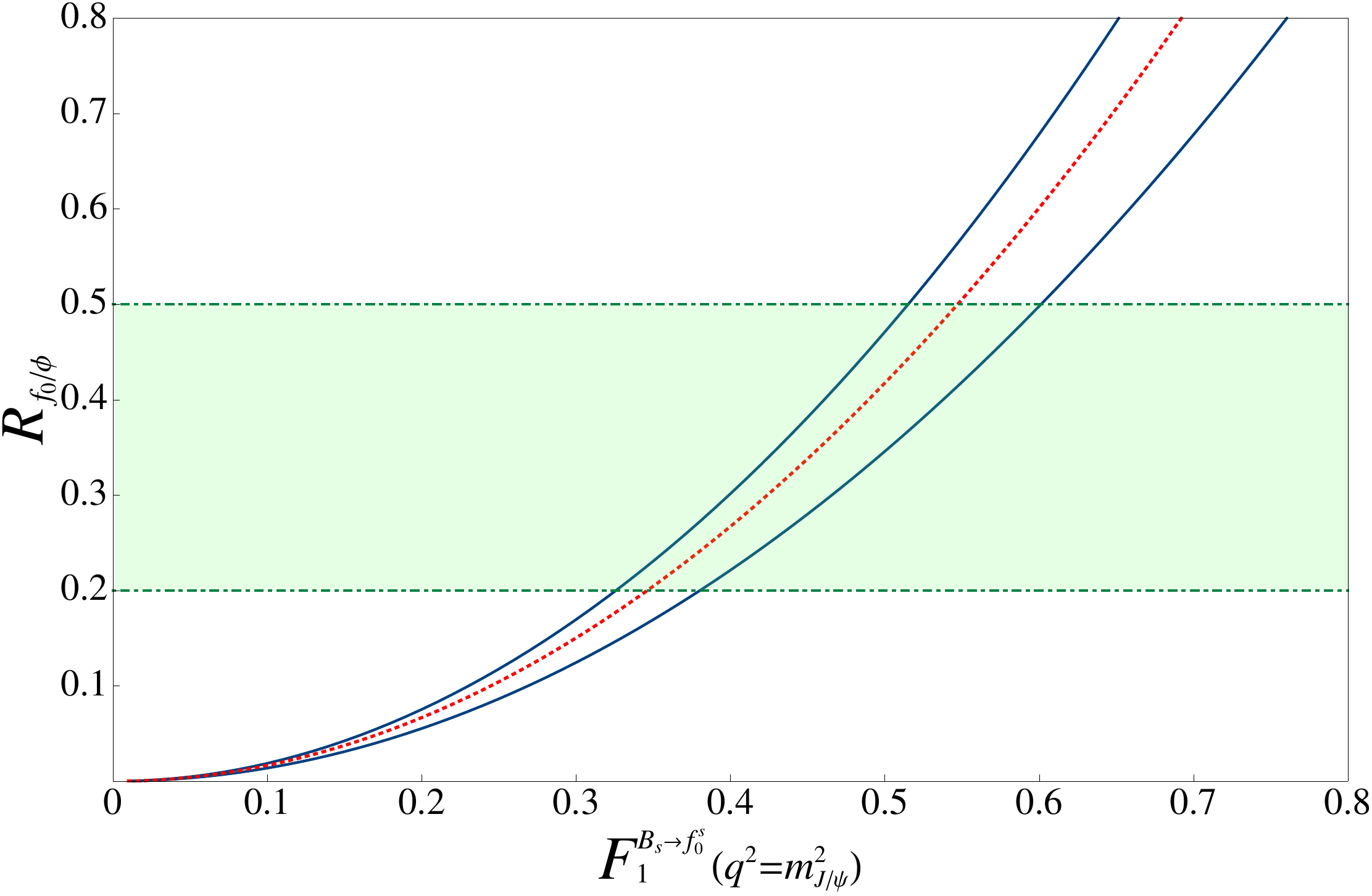}
\caption{The ratio $\mathcal{R}_{f_0/\phi}$ as a function of the transition form factor $F_1^{B_s^0 \to f_0^s}(m_{J/\psi}^2)$. Only tree, vertex and penguin 
contributions ($LOVP$), independent of the decay constants $\fBs$ and $\bar f_{f_0}$, are included in the decay amplitudes. The dotted line corresponds 
to the central value of this ratio while the area between the two solid  lines gives its envelope due to the uncertainties on the decay rates $f_0(980) \to \pi^+ \pi^-$
\cite{ElBennich:2008xy,Ecklund:2009fia}  and $\phi \to K^+ K^-$~\cite{Amsler:2008zzb}. The two horizontal dash-dotted lines delimit the (shaded)  area 
between the experimental predictions  found in Refs.~\cite{Stone:2008ak} and \cite{Ecklund:2009fia}.}
\label{fig1} 
\vspace*{3mm}
\includegraphics*[scale=0.54]{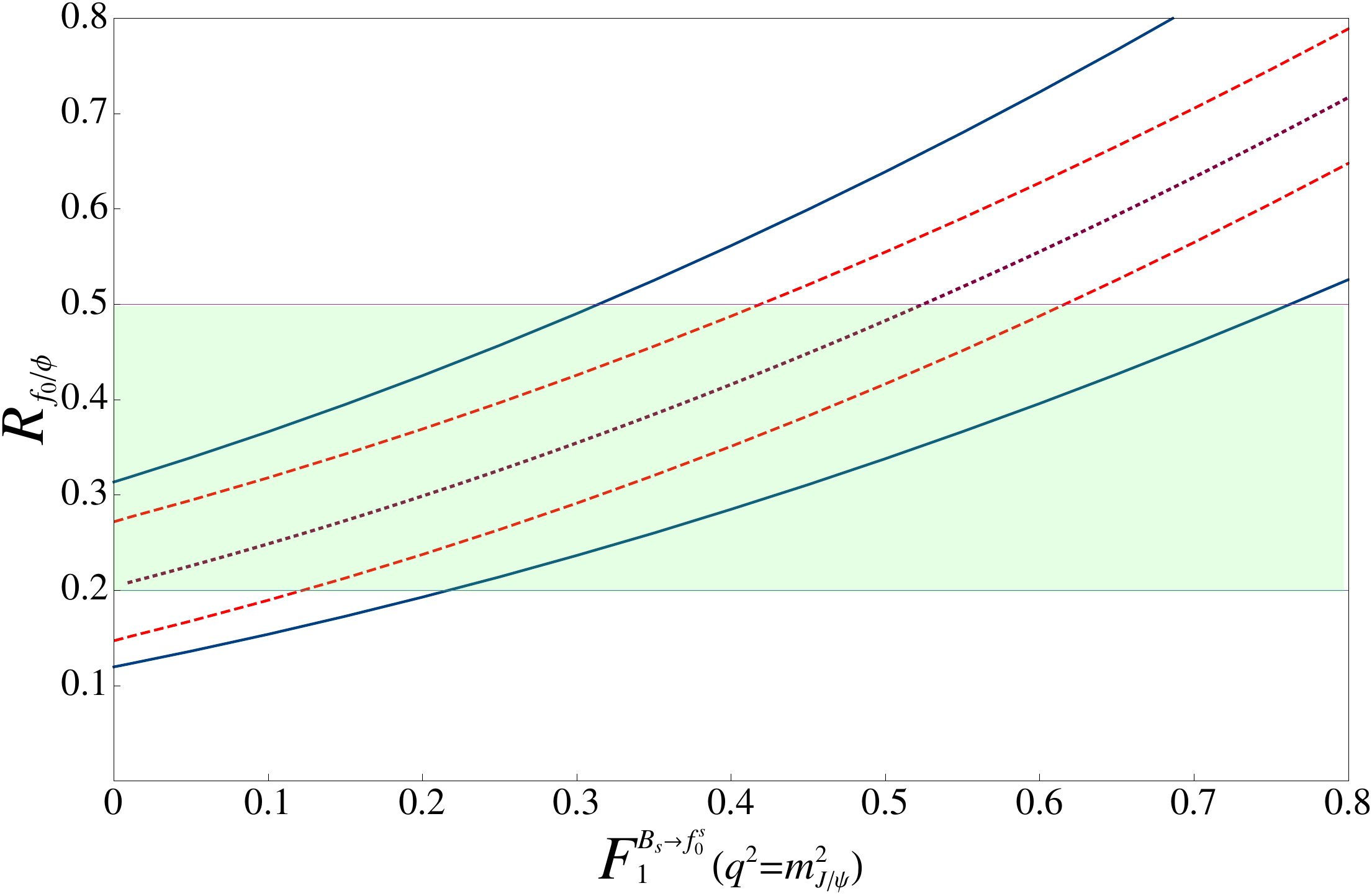}
\caption{The ratio $\mathcal{R}_{f_0/\phi}$ as a function of the transition form factor $F_1^{B_s^0 \to f_0^s}(m_{J/\psi}^2)$ where now the tree, vertex, penguin, 
and hard-scattering contributions ($LOVPH$) are included. The area between the two dashed lines gives the envelope of this ratio when taking into account 
uncertainties on the decay constants ($\fBs =260 \pm 30$ MeV and $\bar f_{f_0}=380 \pm 40$ MeV) while the solid lines include in addition the 
uncertainties on the decay rates $f_0(980) \to \pi^+ \pi^-$~\cite{ElBennich:2008xy, Ecklund:2009fia}  and $\phi \to K^+ K^-$~\cite{Amsler:2008zzb}. 
The single dotted line is our prediction for the central values of the decay constants,  $\fBs =260$~MeV and $\bar f_{f_0}=380$~MeV. 
The horizontal dash-dotted lines correspond to the experimental predictions of Refs.~\cite{Stone:2008ak} and~\cite{Ecklund:2009fia}. }
\label{fig2} 
\end{figure*} }
 
Our prediction for the time-integrated asymmetry $A_{\mathrm{CP}}(B_s \to f_0 J/\psi)$ is about one order of magnitude smaller than the Standard Model 
value, $-2\beta_s  = - 0.036$. We remark that the above numerical values for this $CP$ asymmetry have to be interpreted with care ---  we choose the 
parameters of the full QCDF amplitude in Table~\ref{rhophi} such that the experimental $B_s^0 \to J/\psi \phi$ observables are reproduced. In doing so, we may
deliberately include ``new physics" effects with just the Standard Model amplitude, in particular via the additional amplitudes $\zeta^{(h)}$. 
Moreover, we use the same end-point parameterization, $X_H$,  in both decay channels since the $B_s^0 \to J/\psi f_0$ branching ratio is not experimentally 
known.  This approach seems reasonable, as the physics buried  in these infrared divergences must be similar in both decays. It could also lead to an 
overestimation of the hard-scattering contributions to $B_s^0 \to J/\psi f_0$ as well as of $A_{\mathrm{CP}}(B_s \to f_0 J/\psi)$.

We illustrate the variation of the ratio, $\mathcal{R}_{f_0/\phi}$, by taking into account the uncertainties in the decay constants $\fBs$ and $\bar f_{f_0^s}$ as 
well as those in the decay rates, $\mathcal{B}({f_0(980) \to \pi^+ \pi^- })=0.50^{+0.07}_{-0.09}$~\cite{ElBennich:2008xy,Ecklund:2009fia} and 
$\mathcal{B}({\phi \to K^+ K^- })=0.489\pm 0.005$~\cite{Amsler:2008zzb}.  The results are displayed in Figures~\ref{fig1}--\ref{fig3}.

{\renewcommand\baselinestretch{1.1}
\begin{figure*}[t!]
\includegraphics*[scale=0.54]{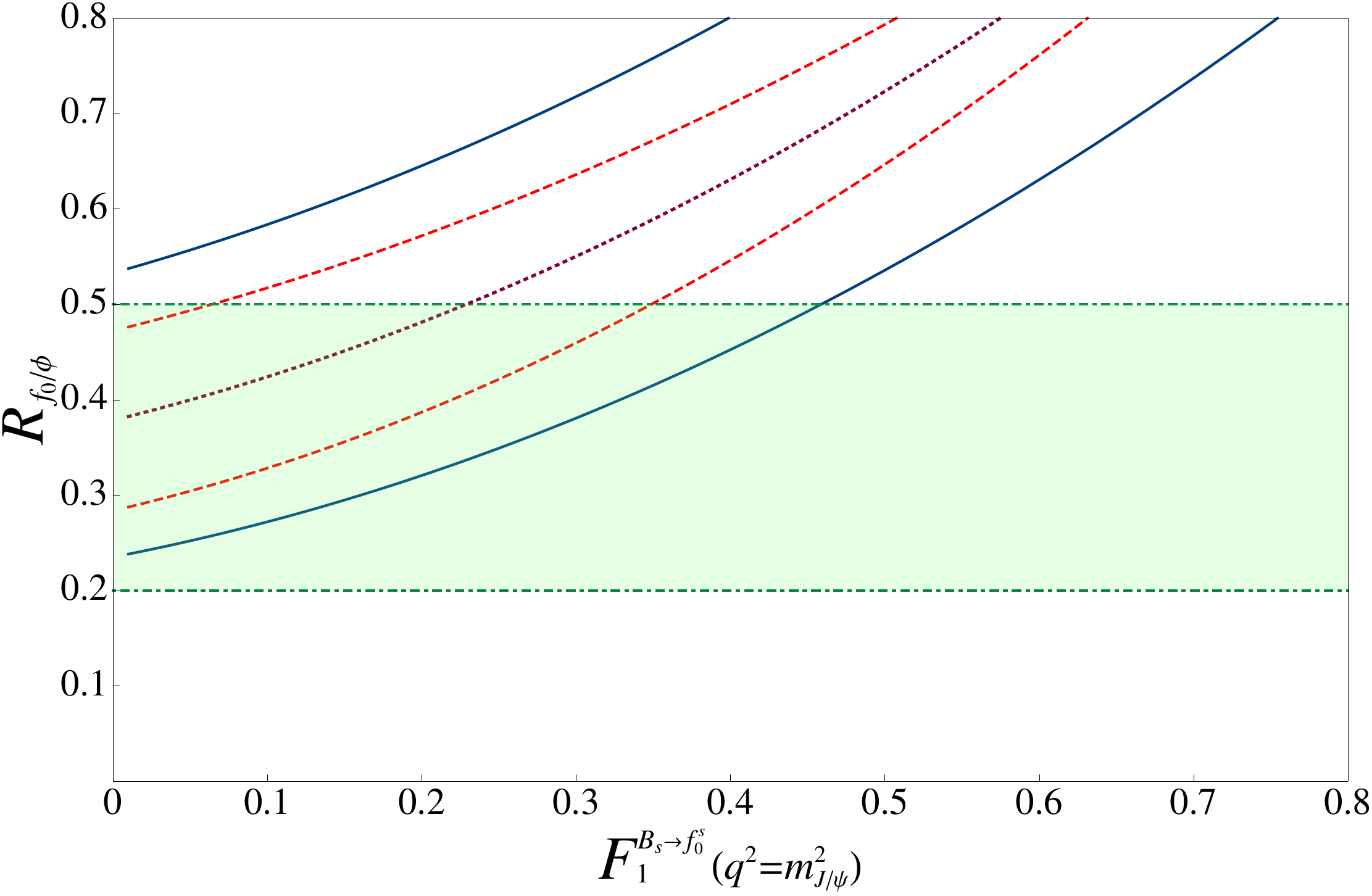}
\caption{Same as in Fig.~\ref{fig2} but including $\zeta^{(h)}$ contributions ($LOVPH+C$ amplitudes).}
\label{fig3}
\end{figure*} }

In Figure~\ref{fig1}, $\mathcal{R}_{f_0/\phi}$ is plotted as a function of $F_1^{B_s^0 \to f_0^s}(m^2_{J/\psi})$ where only the tree amplitude along with vertex 
and penguin corrections are included in both amplitudes, ${\cal A}_{\B_s^0\to \phi J/\psi}^h$ and ${\cal A}_{\B_s^0\to f_0 J/\psi}$. The ratio is plotted with the 
corresponding envelope of $\mathcal{R}_{f_0/\phi}$ due to the uncertainty on the decay rates. In Figure~\ref{fig2}, we augment this amplitude by hard-scattering 
contributions, that is the full QCDF amplitude given in Eq.~\eqref{defa}. Finally, in Figure~\ref{fig3},  $\mathcal{R}_{f_0/\phi}$ is plotted as a function of 
$F_1^{B_s^0 \to f_0^s}(m^2_{J/\psi})$ including hard-scattering corrections {\em and\/} possible ``other" contributions, $\zeta^{(h)}\!$. Although the 
aforementioned uncertainties are depicted in all figures, we stress that those on the decay constants $\fBs$ and $\bar f_{f_0^s}$, where they apply, have more impact 
on the $\mathcal{R}_{f_0/\phi}$ band than the $f_0(980)$ and $\phi$ decay rate incertitudes. The spreading of the curves representing $\mathcal{R}_{f_0/\phi}$ as 
a function of $F_1^{B_s^0 \to f_0^s}(m^2_{J/\psi})$ is larger with respect to the  variation in $\fBs$  than in $\bar f_{f_0^s}$. This points to the necessity of 
having an improved experimental determination of $\fBs$. The upper limit of the envelope is reached only for the largest values of  $\fBs$ and $\bar f_{f_0^s}$
considered here. 

Figure~\ref{fig3} shows that our central-value predictions of $\mathcal{R}_{f_0/\phi}$, in absence of any phenomenological contributions, are within the estimate
by Stone and Zhang~\cite{Stone:2008ak} for most values of the form factor $F_1^{B_s^0 \to f_0^s}(m^2_{J/\psi})$.  However, when the additional amplitudes, 
$\zeta$, are accounted for in the decay amplitudes of Eqs.~\eqref{AVV} and \eqref{ASV},  the ratio $\mathcal{R}_{f_0/\phi}$ exhibits three striking features:
\begin{itemize}	
\item \quad Additional amplitudes, $\zeta$, can play a major role due to their large contributions to both the numerator and denominator of the ratio 
$\mathcal{R}_{f_0/\phi}$, as seen from  the comparison of Figures~\ref{fig2} and~\ref{fig3}.

\item \quad The predicted $\mathcal{R}_{f_0/\phi}$ band overlaps well with the estimates of  Refs.~\cite{Stone:2008ak} and~\cite{Ecklund:2009fia} for  
$F_1^{B_s^0 \to f_0^s}(m^2_{J/\psi})<0.4$; beyond, our predictions are much larger, which
may indicate a larger pollution due to $f_0(980)\to K^+K^-$ if contributions from other than the Standard Model were present.

\item \quad The uncertainties on the $f_0(980)$ and $\phi$ decay rates increase the width of the band considerably, though the main uncertainty stems 
from the decay constants $\fBs$ and $\bar f_{f_0}$.
\end{itemize}

Let us remind that the decay constant $\bar{f}_{f^s_0}$ only enters the hard-scattering and additional phenomenological contributions ($C$) 
to the decay amplitude $B_s^0 \to f_0(980) J/\psi $. If these are turned off, as in Figure~\ref{fig1}, the ratio $R_{f_0/\phi}$ is still significantly above 10\% 
for realistic transition-form factor values. That said, for practical purposes we decide to only consider the more recently obtained decay constants
in Ref.~\cite{Cheng:2005nb} and choose three values within the given errors, $\bar f_{f_0^s} = 340, 380, 420$~MeV. The value $180$~MeV~\cite{DeFazio:2001uc} 
yields too low branching fractions in other decays, for example $B \to f_0(980) \pi,  f_0(980)\rho,  f_0(980) K^{(*)}$. Nevertheless, since we fix the hard-scattering 
parameters, $\rho_H$ and $\phi_H$, only via the decay $B_s^0 \to J/\psi \phi$ and $\bar f_{f_0^s}$ enters the numerator in $\mathcal{R}_{f_0/\phi}$ linearly, the modification 
is straightforward:  $\bar f_{f_0^s} = 180$~MeV is about half the value $\bar f_{f_0^s} = 380$~MeV, therefore the central value of  $\mathcal{R}_{f_0\phi}$ 
in  Figure~\ref{fig2} decreases from $0.42$ to $0.19$ (for $F_0^{B_s^0 \to f_0^s} =0.4$). This is still within the limits predicted by the experimental estimates,
$0.2 \lesssim  \mathcal{R}_{f_0/\phi} \lesssim  0.5$, and implies an $S$-wave pollution.

We infer from our numerical results that $S$-wave kaons or pions under the $\phi$ peak in $B_s^0 \to J/\psi \phi$ are very likely to originate from the similar decay 
$B_s^0 \to J/\psi f_0$. Therefore, the extraction of the mixing phase, $-2 \beta_s$, from $B_s^0 \to  J/\psi \phi$ may well be biased by this $S$-wave effect which
should be taken into account in experimental analyses. In our interpretation of the  full QCDF amplitude, we not only confirm the influence of $S$-wave contamination 
as advocated in Refs.~\cite{Stone:2008ak} and \cite{Ecklund:2009fia} but also find that its effect could be sizable.

\section{Conclusive outlook \label{secseven}}

The ``phase" of $B_s^0 - \bar B_s^0$ mixing, $-2\beta_s$, is thought to be best measured in the golden decay,  $B_s^0 \to J/\psi \phi$, and provides
an opportune place to investigate physics beyond the Standard Model. Several models have been proposed to explain the apparent discrepancy of
the Standard Model prediction for $-2\beta_s$ with recent experiments, in particular exploring the impact of heavy, as of yet undiscovered particles on 
$CP$ violation in weak $B$-meson decays. A general analysis of possible new physics effects in the case of  $B_s^0 - \bar B_s^0$ mixing was recently 
given by \mbox{Chiang} {\em et al.\/}~\cite{Chiang:2009ev}. In there, the authors investigate several beyond Standard Model variations of the $B_s \to J/\psi \phi$ 
decay, such as $Z^{(')}$-mediated Flavor Changing Neutral Currents (FCNC), two Higgs doublets and SUSY, and find that new physics contributions 
may only modestly contribute to the mixing phase. However, it is also concluded, somewhat prematurely, that  the CDF and D$\emptyset$ results are 
clear signs of new physics.

In the present paper, we have taken a different path and studied the contamination of final state $S$-waves kaons in the $B_s^0 \to J/\psi \phi$
channel by those originating from the $f_0(980)$ in the very similar $B_s^0 \to J/\psi f_0(980)$ decay. We find that this effect is strong enough
already for amplitudes including leading order, vertex and penguin corrections to create a real bias in the determination of $-2 \beta_s$. 

Of course, we are aware that the phenomenological endpoint parametrization of $\alphas$ corrections in the amplitudes $ H_n(M_1 J/\psi)$ and $ H_n^h(M_1 J/\psi)$
can cloud possible new physics contributions alongside the $\zeta^{(h)}$ contributions. In this case, we suppose that any new effects should be of comparable 
magnitude in $B_s^0 \to J/\psi \phi$ and $B_s^0 \to J/\psi f_0(980)$. Therefore,  the $S$-wave contamination would be on the upper side of the estimate we 
propound and future analyses of the mixing angle in $B_s$ decays should be concerned with this effect.

\begin{acknowledgments}
The authors are obliged to Sheldon Stone for pointing out the likely $S$-wave effects on the mixing angle which stimulated this work and critical
remarks on the manuscript. 
We thank Martin Beneke for useful comments on mass corrections in transition form factors between two mesons. 
We are very grateful to Ying Li for drawing our attention to the particular quark topology of the annihilation diagrams.
B.~E. wishes to acknowledge the hospitality of the {\it Laboratoire de Physique Nucl\'eaire et Hautes Energies\/} in Paris and of Jo\~ao Pacheco B.~C. 
de Melo and Victo dos Santos Filho at Universidade Cruzeiro do Sul, S\~ao Paulo, during the final stage of the writing. This work was partially funded 
by the Department of Energy, Office of Nuclear Physics, under contract number DE-AC02-06CH11357.
\end{acknowledgments}



\end{document}